\documentclass[10pt,reqno,oneside]{amsart}
\usepackage{verbatim,amsmath,amssymb,cite,epsfig,subfigure,color,hyperref,ulem}

\numberwithin{equation}{section}

\begin{document}

\title[Possible Implications of Self-Similarity for Tornadogenesis and Maintenance]
{Possible Implications of Self-Similarity for Tornadogenesis and Maintenance}

\author{Pavel B\v{e}l\'{\i}k}
\author{Brittany Dahl}
\author{Douglas Dokken}
\author{Corey K.~Potvin}
\author{Kurt Scholz}
\author{Mikhail Shvartsman}

\address{P.~B\v{e}l\'{\i}k\\
Department of Mathematics, Statistics, and Computer Science\\
Augsburg University\\
2211 Riverside Avenue\\
Minneapolis, MN 55454\\
U.S.A.}
\email{belik@augsburg.edu}

\address{B.~Dahl\\
School of Meteorology\\
University of Oklahoma\\
Norman, OK\\
U.S.A.}
\email{}

\address{C.~Potvin\\
Cooperative Institute for Mesoscale Meteorological Studies, and School of Meteorology, University of Oklahoma, and NOAA/OAR/National Severe Storms Laboratory\\
Norman, OK\\
U.S.A.}
\email{}

\address{D.~Dokken\\
Department of Mathematics\\
University of St.~Thomas\\
2115 Summit Ave.\\
St.~Paul, MN 55105\\
U.S.A.}
\email{dpdokken@stthomas.edu}

\address{K.~Scholz\\
Department of Mathematics\\
University of St.~Thomas\\
2115 Summit Ave.\\
St.~Paul, MN 55105\\
U.S.A.}
\email{k9scholz@stthomas.edu}

\address{M.~Shvartsman\\
Department of Mathematics\\
University of St.~Thomas\\
2115 Summit Ave.\\
St.~Paul, MN 55105\\
U.S.A.}
\email{mmshvartsman@stthomas.edu}

\thanks{
  The authors would like to acknowledge Amy McGovern from the University of Oklahoma for contributing the data used in Section~\ref{sec:cm1}. They would also like to acknowledge the constructive feedback provided by Cheri Shakiban from the University of St.~Thomas. Funding for Potvin was provided by the NOAA/Office of Oceanic and Atmospheric Research under NOAA--University of Oklahoma Cooperative Agreement \#NA11OAR4320072, U.S.~Department of Commerce. Funding for Dahl was provided by NSF/IIS grant 0746816.
}



\date{\today}

\begin{abstract}
  Self-similarity in tornadic and some non-tornadic supercell flows is studied and power laws relating various quantities in such flows are demonstrated. Magnitudes of the exponents in these power laws are related to the intensity of the corresponding flow and thus the severity of the supercell storm. The features studied in this paper include the vertical vorticity and pseudovorticity, both obtained from radar observations and from numerical simulations, the tangential velocity, and the energy spectrum as a function of the wave number. Connections to fractals are highlighted and discussed.
\end{abstract}

\allowdisplaybreaks
\thispagestyle{empty}
\maketitle

\section{Introduction}
\label{sec:intro}
Power laws with a particular scaling exponent arise when a phenomenon ``repeats itself on changing scales'' \cite{mandelbrot83,barnsley88,barenblatt96,barenblatt03}. This property is called self-similarity. We propose that the study of strong atmospheric vortices requires further exploration of their self-similarity, since self-similarity can point to important properties of the underlying dynamics. As will be shown, tornadoes appear to exhibit local self-similarity suggesting fractal phenomena and that will be the focus of this paper. A possible way tornadoes and mesocyclones might acquire self-similarity is through a vortex sheet roll-up; this process would give rise to the hypothesized vorticity and velocity power laws discussed below.

Self-similarity can manifest itself in several ways in atmospheric flows. One such manifestation is scale-invariance of some characteristic of the flow, which may be demonstrated by the existence of a power law for the characteristic. Examples include the scenarios discussed below, where power laws for vorticity/pseudovorticity or velocity are hypothesized \cite{wurmangill00,wurmanalexander05,cai05}.

Self-similarity is also observed in \cite{belik14}, which revisits Serrin's ``swirling vortex'' model \cite{serrin} and investigates solutions to the Navier--Stokes and Euler equations in spherical coordinates with the velocity, $\bf v$, satisfying the power law $|{\bf v}|\propto r^{-\alpha}$, where $r$ is the distance from the vertical coordinate axis and $\alpha$ is not necessarily equal to $1$. The streamlines and other physical quantities of the modeled vortices, such as isobars, exhibit self-similarity.

\begin{figure}
  \begin{center}
    \includegraphics[height=0.4\textheight]{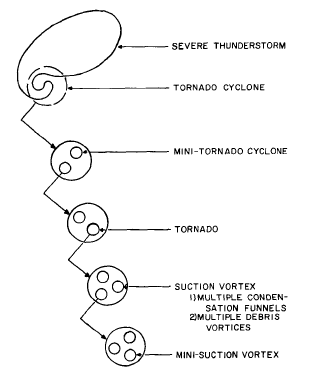}
  \end{center}
  \caption{Hierarchy of known vortex scales in tornadic supercells; \copyright~AMS, \cite{church77}.}
  \label{fig:hier-vort}
\end{figure}

\begin{figure}
  \begin{center}
    \includegraphics[width=0.89\textwidth]{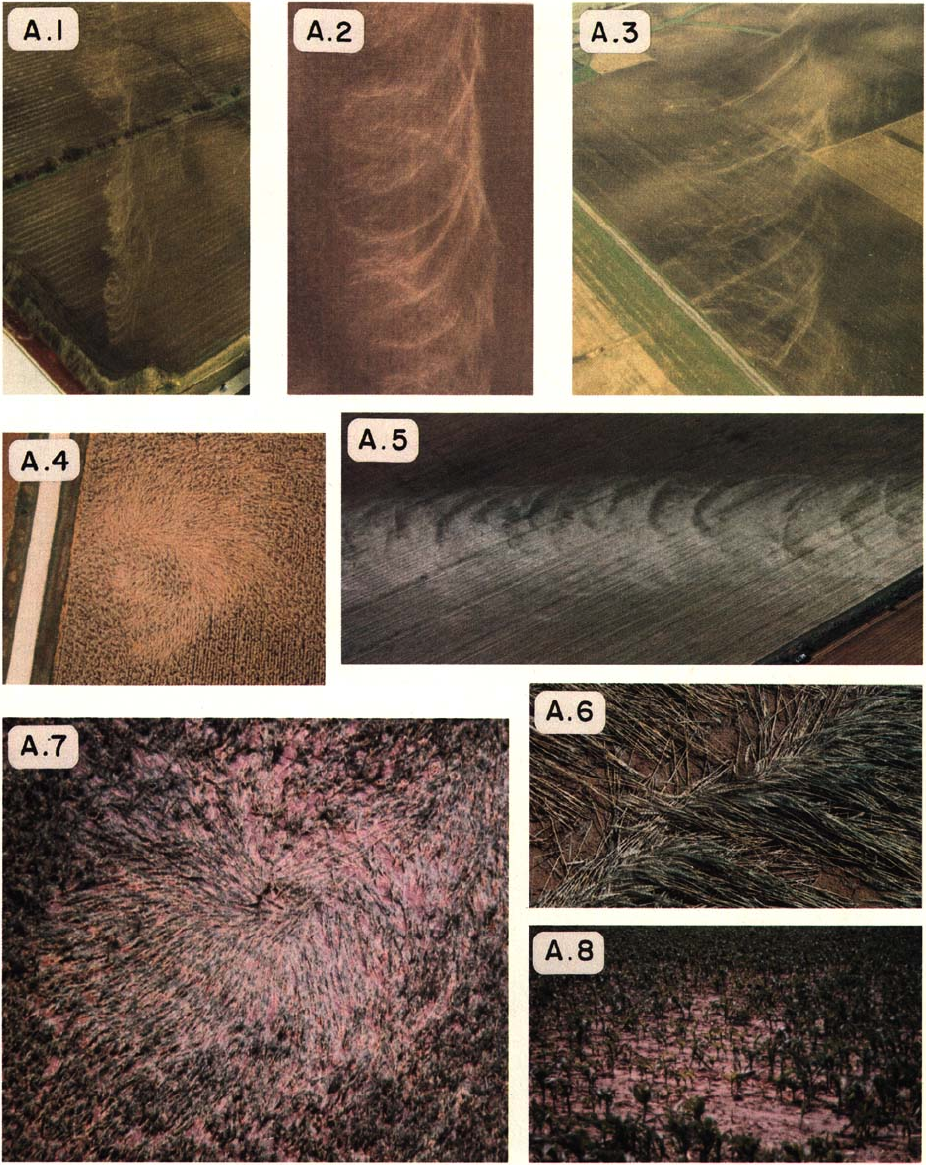}
  \end{center}
  \caption{Examples of ground marks left behind by suction vortices embedded inside tornadoes. Locations and dates of occurrences are: (A.1) Decatur, Illinois tornado 3 April 1974; (A.2) Magnet, Nebraska tornado, 6 May 1975; (A.3) Homer Lake, Indiana tornado, 3 April 1974; (A.4) Dubuque, Iowa tornado, 28 September 1972; (A.5) and (A.6) Pearsall, Texas tornado, 15 April 1973; (A.7) Mattoon Lake, Illinois tornado, 21 August 1977; (A.8) Grand Island, Nebraska tornado, 3 June 1980. \copyright~AMS, \cite{fujita81}.}
  \label{fig:suctionspots}
\end{figure}

Geometric self-similarity in tornadoes manifesting itself over several scales is not a new concept \cite{church77}. Figure~\ref{fig:hier-vort}, taken from \cite{church77}, illustrates a hierarchy of known vortex scales in tornadic supercells. Phenomena occurring at all of these scales may not be easy to observe, but videos of recent large tornadoes show subvortices of subvortices within tornadoes \cite{wadena10,elrenovideo13}. These sub-subvortices, called suction vortices or suction spots, are short lived and very intense, and their existence is often confirmed by studying the track of a tornado afterwards. See Figure~\ref{fig:suctionspots} for examples of such evidence \cite{fujita81}.

Some observed tracks left by the suction vortices within a tornado are as narrow as $30$ cm. Some of these paths appear to originate outside the tornado and intensify as they move into the tornado. We identify these vortices as supercritical in the sense of \cite{fiedlerrotunno86}. Analysis of the work in \cite{barcilon67,Burggraf77,benjamin62} suggests that the supercritical vortex below a vortex breakdown has its volume and its length decrease as the energy of the supercritical vortex increases. This suggests that the entropy (randomness of the vortex) is decreasing when the energy is increased \cite{vortexgas17}. Hence the inverse temperature, which is the rate of change of the entropy with respect to the energy of the vortex, is negative. This temperature has to be considered in the statistical mechanics sense and is not related to the molecular temperature of the atmosphere. Such vortices would be barotropic, however their origin could very well be baroclinic. Recent results suggest that vorticity is produced baroclinically in the rear-flank downdraft and then descends to the surface, where it is tilted into the vertical, contributing to tornadogenesis. Even more recently, simulations show vortices produced in the forward flank region contributing to tornadogenesis and maintenance \cite{orf17}. Once these vortices come into contact with the surface, and the stretching and surface friction related swirl (boundary layer effects) are in the appropriate ratio, then by analogy with the work in \cite{fiedlerrotunno86} the vortex would have negative temperature and the vortex would now be barotropic \cite{davies15}. 

\begin{figure}
  \begin{center}
    \includegraphics[width=\textwidth]{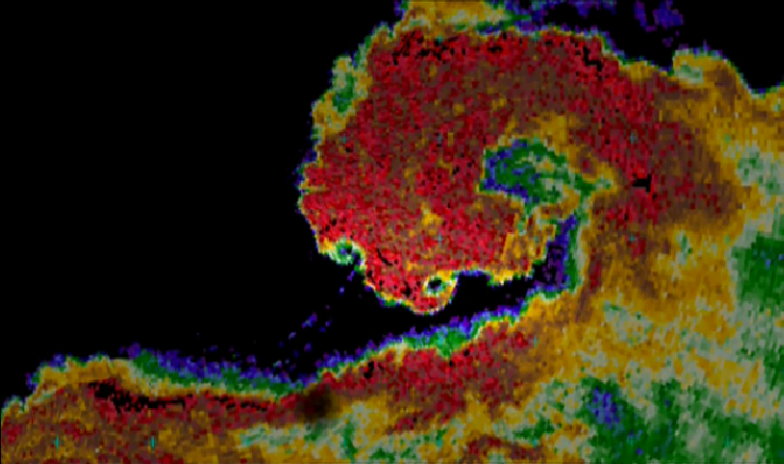}
  \end{center}
  \caption{A reflectivity image of a tornado showing self-similarity and a possible fractal structure; \copyright~Joshua Wurman, \cite{nova04}.}
  \label{fig:tornado_fractal}
\end{figure}

Geometric self-similarity is occasionally seen in high-resolution numerical simulations of tornadic supercells \cite{lewellenssykes97,adlerman00,eyink13} and also in Doppler radar and reflectivity observations \cite{bluestein00b,nova04}. As an example, in the reflectivity image in Figure \ref{fig:tornado_fractal} we can see self-similarity on two different scales demonstrating itself as ``hooks on a hook.'' This is likely due to the existence of subvortices within the larger vortex. High-quality video recordings of some recent tornadoes depict mini suction vortices (subvortices of suction vortices), confirming the smallest scale of the hierarchy in Figure \ref{fig:hier-vort} \cite{wadena10,elrenovideo13}.

\begin{figure}
  \begin{center}
    \includegraphics[width=\textwidth]{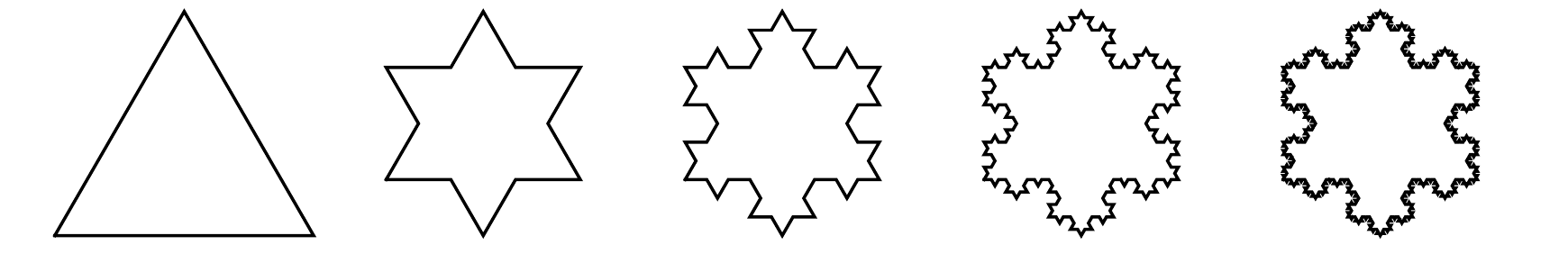}
  \end{center}
  \caption{Construction of the Koch snowflake starting from an equilateral triangle and showing the first four steps. The limiting fractal object is known as the Koch snowflake and it has a fractal dimension of $\log{4}/\log{3}\approx1.26186$.}
  \label{fig:snowflake}
\end{figure}

Fractals are mathematical objects useful as idealizations of structures and phenomena in which features and patterns repeat on progressively smaller and smaller scales \cite{mandelbrot83}. Such structures exhibit geometrical complexity that can be, in a simplified way, captured by a fractal dimension of the object, a number that describes how the fractal pattern changes with scale. For example, the fractal dimension of the well-known Koch snowflake shown in Figure~\ref{fig:snowflake} is $\log{4}/\log{3}\approx1.26186$, which indicates that the object has enough ``turns'' to not be a $1$-dimensional curve, but is far from filling up a planar region of dimension $2$. An early discussion of fractals in fluid mechanics can be found in \cite{turcotte88}.

Chorin discusses quantities with fractal dimensions in his study of turbulent flows \cite{chorin}. Through numerical experiments he finds fractal dimensions of the axes of vortices he studies to be related to the ``temperature'' of the vortex. ``Hot'' negative-temperature vortices have a smooth axis, while at temperatures of positive or negative infinity the vortex has a fractal axis. Under these hypotheses one should expect that high-energy vortices entering the tornado acquire fractal axes upon being stretched and kinked up (transition from negative-temperature vortices to infinite-temperature vortices). One would expect a mixture of fractal dimensions for these axes in the turbulent region surrounding the solid body tornado core.

In their study of the effect of rotation and helicity on self-similarity, Pouquet et al.\ state that ``when comparing numerical simulations, it was found that two runs at similar Rossby numbers and at similar times (albeit at different Reynolds numbers) display self-similar behavior or decreased intermittency depending on whether the flow had helicity or not'' \cite{pouquet10}. That tornadoes form in helical environments may largely account for the degree of self-similarity that is often observed in them (i.e., the presence of suction vortices), and suggests self-similarity may extend to smaller scales than currently known. We propose that such self-similarity can arise within persistent vortex sheets along the rear flank and forward flank downdraft gust fronts of tornadic supercells. In the proposed scenario, a sequence of vortex roll-ups occurs, with each new generation of vortices forming from previous-generation vortices wrapping around each other, ultimately resulting in vortices with roughly fractal cross sections (geometric self-similarity).

The remainder of this paper is organized as follows. In Section~\ref{sec:wurman} we discuss both observed and studied power laws in the tangential velocity of tornadoes as a function of the radial distance from the axis of the vortex. In Section~\ref{sec:pseudovorticity} we briefly review the work in \cite{cai05} and discuss power laws in the vertical vorticity and pseudovorticity as a function of scale for some tornadic and nontornadic mesocyclone data obtained from Doppler and dual Doppler data. In Section~\ref{sec:cm1} we discuss the results of a supercell thunderstorm simulation using the Bryan Cloud Model 1 and observe an agreement between the work in \cite{cai05} and a resulting power law for vorticity over multiple scales. In Section \ref{sec:energy_spectrum} we use the vortex gas model developed in our related paper \cite{vortexgas17}, and use a modified argument due to Chorin to show that an increase in fractal dimension of the cross section of a negative temperature vortex corresponds to an increase in energy at large scales.  
Finally, Section~\ref{sec:conclusions} offers conclusions and describes future work.

\section{Power laws in the tangential velocity of tornadoes}
\label{sec:wurman}
While it has been known since at least the 1950s that in tropical cyclones the tangential component of the velocity, $v$, exhibits decay proportional to $r^{-\alpha}$ with $0<\alpha<1$, where $r$ is the radial distance from the center of the cyclone \cite{mallenmontgomerywang2005}, it has been only much more recently that similar power laws have been observed for the tangential component of the velocity in tornadoes \cite{wurmangill00,wurman02,wurmanalexander05,kosibatrappwurman08}. Such power laws could perhaps have been anticipated based on the results obtained earlier in a vortex simulator \cite{lund-snow93}, in which a power law of the form $r^{-0.63}$ has been found for a much smaller physical scale and very different Reynolds number. In fact, recent theoretical results suggest that power laws with similar exponents hold across a whole range of scales, ranging from a bathtub vortex, through dust devils and firewhirls, to tornadoes and tropical cyclones \cite{klimenko14}.

We now briefly review some of the observations made by Wurman and his collaborators based on mobile Doppler radar data analyses \cite{wurmangill00,wurman02,wurmanalexander05,kosibatrappwurman08}. In these papers, the tangential winds outside the tornado core roughly fit the {\it modified Rankine} vortex model, in which the core is modeled as a solid-core rotation with the mean tangential velocity depending linearly on radius (i.e., $v\propto r$) and outside of it decaying proportionally to $r^{-\alpha}$ (i.e., $v\propto r^{-\alpha}$). These results are summarized in Table~\ref{tab:alpha}.
\begin{table}
  \begin{tabular}{cc}
    Tornadic storm & Value of $\alpha$ \\
    \hline
    F2--F4, Dimmit, TX, June 2, 1995 & $0.5$-$0.7$ \\
    F4, Spencer, SC, May 30, 1998 & $\approx0.67$ \\
    F4, Mulhall, OK, May 3, 1999 & $0.5$-$0.6$ \\
    F0--F4, Harper County, KS, May 12, 2004 & $0.26\rightarrow0.61$
  \end{tabular}
  \caption{Summary of values of the power law exponents $\alpha$ in $v\propto r^{-\alpha}$ for various tornadic storms mentioned in the text.}
  \label{tab:alpha}
\end{table}

\begin{figure}
  \begin{center}
    \includegraphics[width=\textwidth]{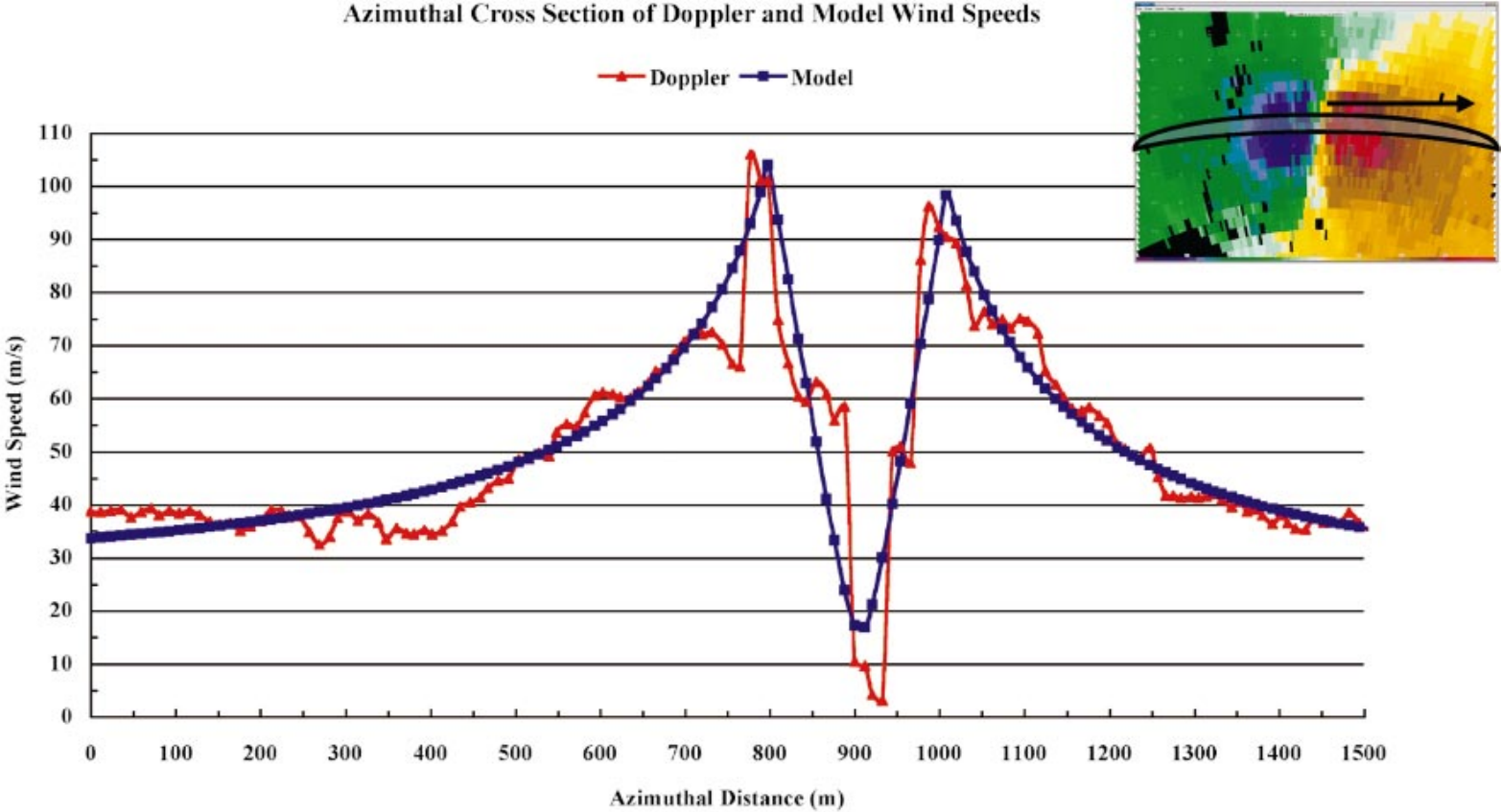}
  \end{center}
  \caption{A profile of measured tangential winds (red) together with a fitted modified Rankine vortex model (blue) in a tornado that occurred in Spencer, SD on May 30, 1998. \copyright~AMS \cite{wurmanalexander05}.}
  \label{fig:wurman-alexander-decay}
\end{figure}

The results of the analysis of data collected from an F2--F4 tornado that occurred in Dimmit, Texas on June 2, 1995 indicate that the exponent $\alpha$ was in the range $0.5$ to $0.7$ \cite{wurmangill00}.
The results of the analysis of data collected from an F4 tornado that occurred in Spencer, South Dakota on May 30, 1998 indicate that the exponent $\alpha\approx0.67$ provides the best fit for the collected data \cite{wurmanalexander05}. The measurements together with the fitted modified Rankine vortex model are shown in Figure~\ref{fig:wurman-alexander-decay}.
The results of the analysis of data collected from an F4 tornado that occurred near Mulhall, Oklahoma on May 3, 1999 indicate that over the $8$-minute observation period the exponent $\alpha$ was in the range $0.5$ to $0.6$, although ``the decay rate was sometimes closer to $V\propto R^{-1}$'' \cite{wurman02}.
The results of the analysis of data collected from an F0--F4 tornado that occurred in Harper County, Kansas on May 12, 2004 indicate that over the $13$-minute observation period the tornado intensified while at the same time $\alpha$ increased from $0.26$ to $0.61$ \cite{kosibatrappwurman08}.

These results strongly indicate that established tornadoes exhibit a power-law decay in the tangential velocity and that the magnitude of the exponent correlates with the severity of the tornado. This is consistent with some of the authors' work \cite{belik14}, in which Serrin's ``swirling vortex'' model \cite{serrin} is revisited and solutions to the Navier--Stokes and Euler equations are sought for which velocity is proportional to $r^{-\alpha}$ with a general positive $\alpha$. It is shown that only solutions with $0<\alpha\le1$ are physically reasonable, and that more violent storms would correspond to larger values of $\alpha$ and less violent ones to smaller values of $\alpha$.

Some justifications and explanations for the power laws in tornadoes can be inferred from existing literature. For example, in the case of the Mulhall tornado, evidence is provided supporting both the creation of vortices inside the tornado and outside of it resulting in flows potentially enhancing the tornado's strength \cite{wurman02}. The vortices apparently originating in the vortex core appear to make a partial revolution about the ambient tornado vortex and then dissipate. These secondary vortices have a different velocity and shear profile than the parent tornadoes. They appear to be single celled with extreme values of shear and extreme transient updrafts. The tornadoes themselves appear to have a two-cell structure and a modified Rankine combined profile described above. However, in an extreme case, a power law $v\propto r^{-1}$ is found on one side of an intense tornado. Such a power law would be consistent with no vorticity outside the tornado core on that side. On the other side of the tornado, the power law is found to be $v\propto r^{-\alpha}$ with $0.5\le\alpha\leq0.6$, which is consistent with vorticity being advected into the tornado, possibly along a vortex sheet spiraling into it \cite{orf17}. Conceivably, a distribution of vorticity outside of the tornado core can, in general, be such as to create an average tangential velocity decay following a particular power law, whether this occurs uniformly around the tornado or inhomogeneously as in the case of the Mulhall tornado. Additional comments about inhomogeneous properties of such flows will be provided in section \ref{sec:energy_spectrum}.

\section{Power laws in vorticity and pseudovorticity}
\label{sec:pseudovorticity}
The evidence of power laws for tangential velocity discussed in the previous section suggests that similar power laws should also exist for vertical vorticity. Several tornadic and nontornadic mesocyclones have been studied in \cite{cai05} from the point of view of vertical vorticity and pseudovorticity, and power laws have been discovered to hold over several magnitudes of the length scale.

Specifically, in \cite{cai05} the vertical vorticity was obtained from dual-Doppler radar data. The velocity data were interpolated in a standard way \cite{wakimotoliucai98,wakimotocai00,dowell02a,dowell02b} to grids with horizontal spacing ranging from $\varepsilon=300$ m to $\varepsilon=9600$ m and maximum vertical vorticities, $\zeta_{\text{max}}$, were computed. A strong linear correlation with a linear correlation coefficient close to $1$ was found between $\log{\zeta_{\text{max}}}$ and $\log{\varepsilon}$ and the best fit lines were termed {\it vorticity lines}. If we denote the (negative) slopes of the vorticity lines by $-\beta$, then this means that
\begin{equation}
  \label{eq:zeta_power_law}
  \zeta_{\text{max}}
  \propto
  \varepsilon^{-\beta},
\end{equation}
and this, in turn, is implied by a power law for the tangential velocity discussed in section \ref{sec:wurman}. In particular, under the assumption of axisymmetry, $v\propto r^{-\alpha}$ with $-\alpha=-\beta+1$ for $r$ between $300$ m and $9600$ m would also imply that $\zeta\propto r^{-\beta}$, which, in turn, implies \eqref{eq:zeta_power_law}. The vorticity lines' slopes, $-\beta$, changed over the lifetime of the storms and generally were larger in magnitude for tornadic storms and smaller for nontornadic storms.

A similar approach was applied to pseudovorticity, defined as $\zeta_\text{pv}=\Delta V/L$, where $\Delta V=|(V_r)_\text{max}-(V_r)_\text{min}|$ is the difference between the maximum and minimum radial velocity of the mesocyclone and $L$ is the distance between them. The pseudovorticity can be easily obtained from a single-Doppler velocity field and thus this approach holds practical advantages over the vorticity approach. Analogous to the vorticity results, the {\it pseudovorticity lines}, or the lines of best fit for $\log{\zeta_{\text{pv}}}$ vs.~$\log{\varepsilon}$, were found to fit the data very well (with correlation coefficients between $0.82$ for a nontornadic storm and $0.98$ for an F3 tornadic storm). This then again means that for some $\beta>0$
\begin{equation}
  \label{eq:cai_power_law}
  \zeta_{\text{pv}}
  \propto
  \varepsilon^{-\beta}.
\end{equation}
It was again observed that steeper pseudovorticity lines corresponded to stronger storms, and an empirical threshold between nontornadic and tornadic storms was determined to be the slope of approximately $-1.6$. The larger (in absolute value) vorticity/velocity power law exponents found in strongly tornadic mesocyclones are consistent with the observation that ``parcels that nearly conserve angular momentum penetrate closer to the central axis of the tornadic mesocyclones, resulting in large tangential velocities'' \cite{trapp99}.

In both vorticity and pseudovorticity approaches, the calculated regression lines strongly fit the data over scales between that of the mesocyclone core and that of the ``edge'' of the mesocyclonic tangential flow, indicating that a (pseudo)vorti\-ci\-ty vs.~scale power law is valid over those scales. The exponent $\beta$ in \eqref{eq:cai_power_law} may be interpreted as a fractal dimension associated with the vortex \cite{cai05}. The vorticity power law, if valid, may extend to smaller (including tornadic) scales, but this could not be determined given the limited resolution of the radar observations in \cite{cai05}. As noted in \cite{cai05} and discussed in the previous section, hurricanes exhibit a similar velocity power law and exponent outside their eyewall \cite{mallenmontgomerywang2005,miller67}. Hence roughly the same vorticity power law seems to apply over a range of atmospheric vortex scales.

\section{Vorticity lines computed from a numerical simulation}
\label{sec:cm1}
\begin{figure}
  \begin{center}
    \includegraphics[width=0.75\textwidth]{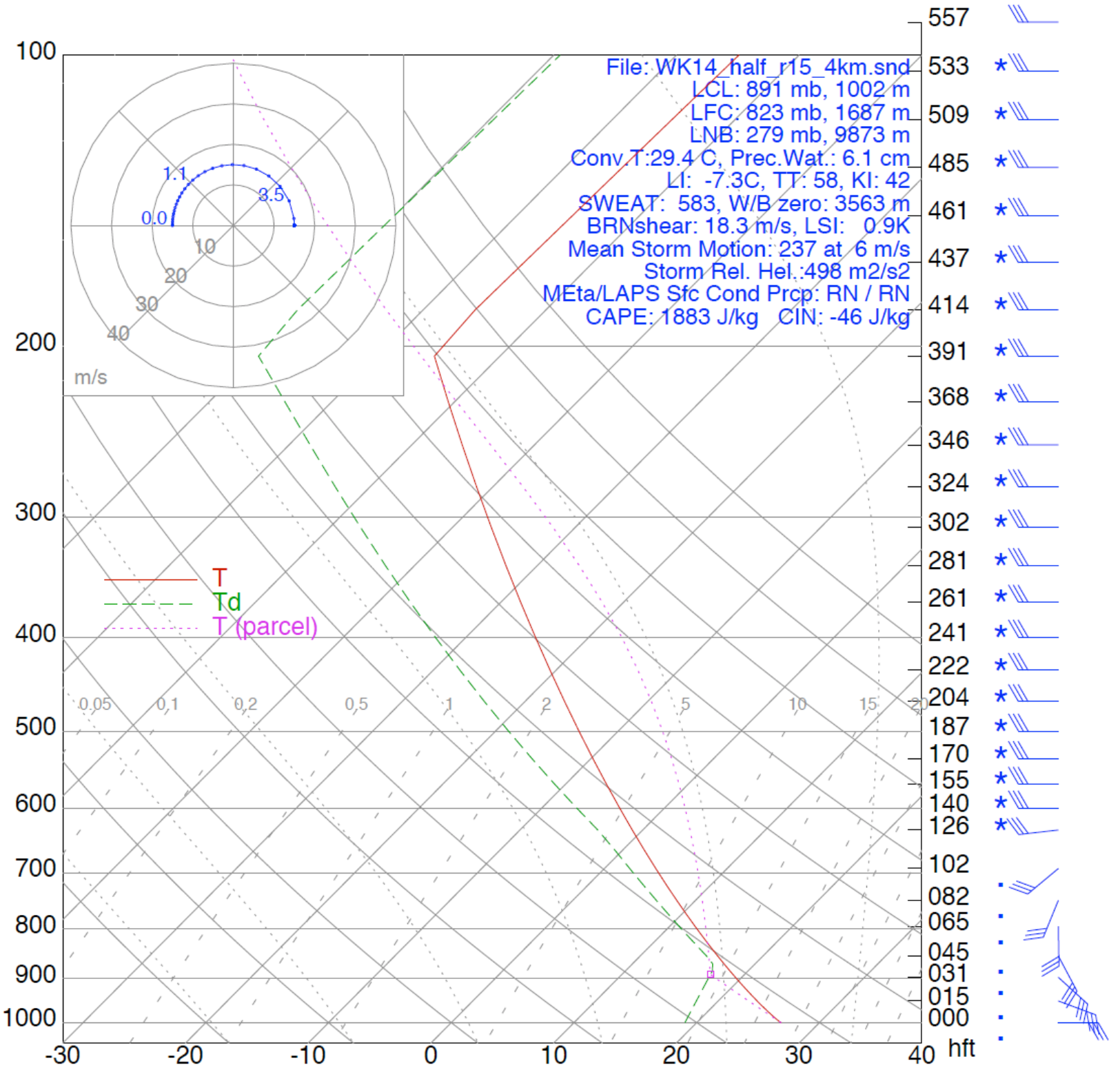}
  \end{center}
  \caption{Base state for a CM1 simulation. Wind barbs are in knots. The hodograph in the upper left corner is in m s$^{-1}$, with marked heights in km.}
  \label{skewt-ok-nssl}
\end{figure}

\begin{figure}
  \begin{center}
  \subfigure[Vertical vorticity field (s$^{-1}$; shaded), horizontal wind vectors (arrows), and $0$ dBZ and $30$ dBZ reflectivity contours at $\sim0.5$ km AGL, $t=158$ min.]
  {
    \hskip 0.4cm
    \includegraphics[height=0.395\textheight]{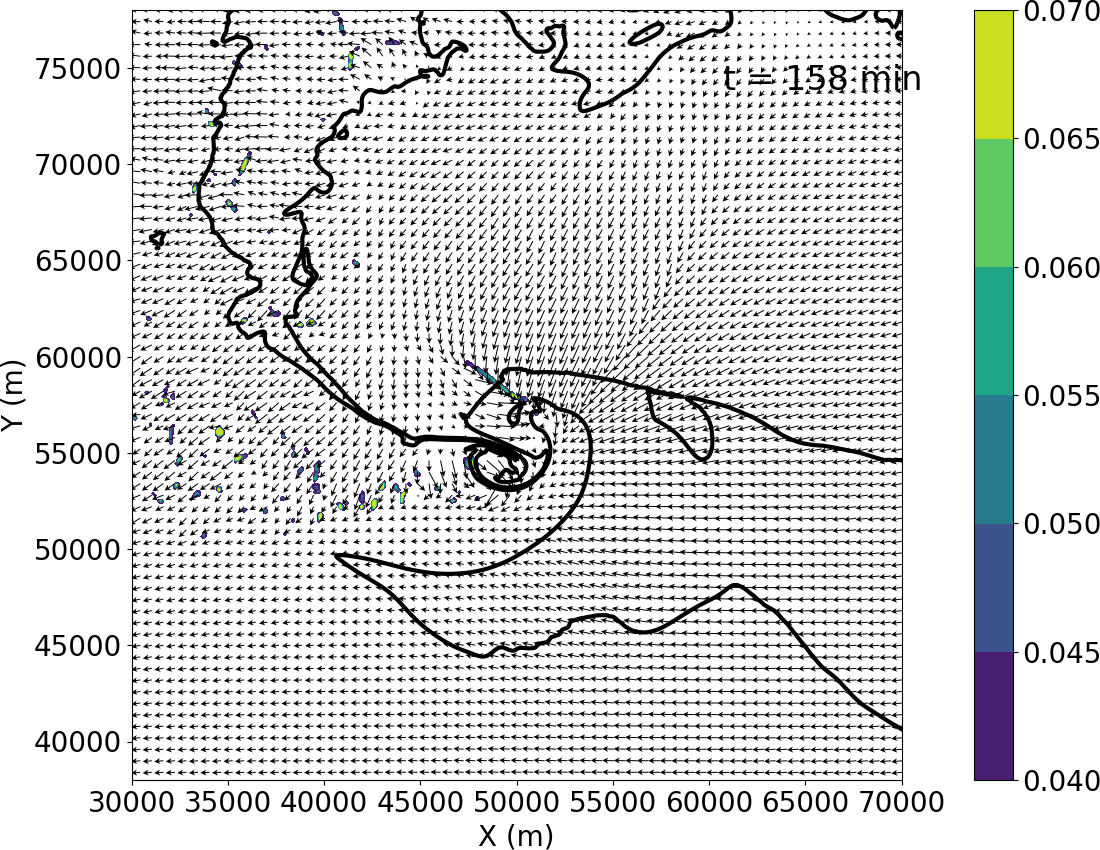}
    \label{fig:158_zoomed-out}
  }
  \subfigure[A simulated reflectivity field (dBZ) and horizontal wind vectors (arrows) at $\sim 0.5$ km AGL and $t=164$ min, near the time of tornadogenesis.]
  {
    \includegraphics[height=0.395\textheight]{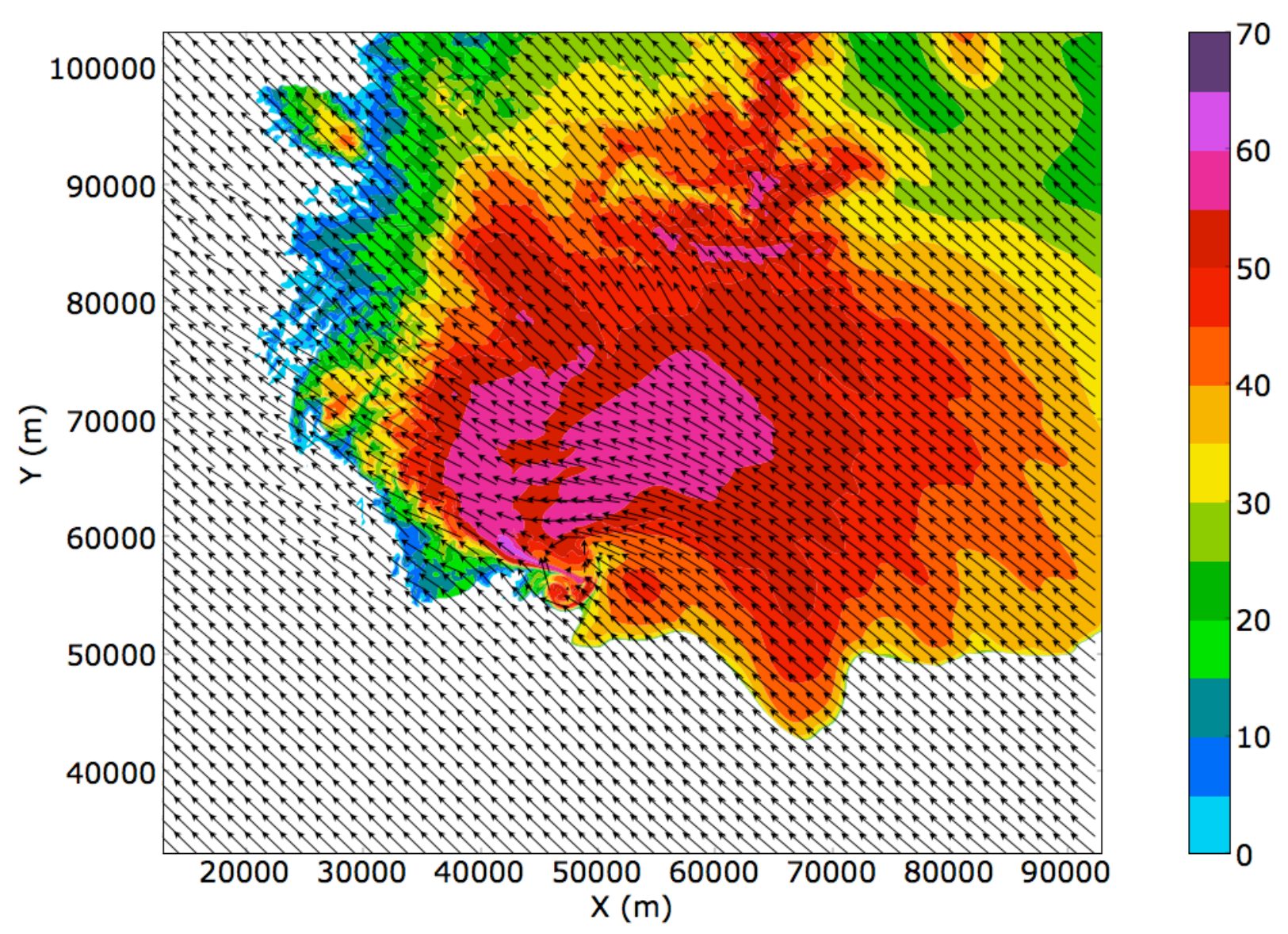}
    \label{fig:simulated_fields}
  }
  \end{center}
  \caption[]{Results from a CM1 simulation exhibiting a cyclonic--anticyclonic vorticity couplet near $x=50$ km and $y=55$ km. \subref{fig:158_zoomed-out} A vortex sheet at $t=158$ min associated to an internal rear flank downdraft just north of the couplet; \subref{fig:simulated_fields} a hook echo reflectivity signature at $t=164$ min.}
  \label{fig:simulated_features}
\end{figure}

\begin{figure}
  \begin{center}
    \includegraphics[width=0.77\textwidth]{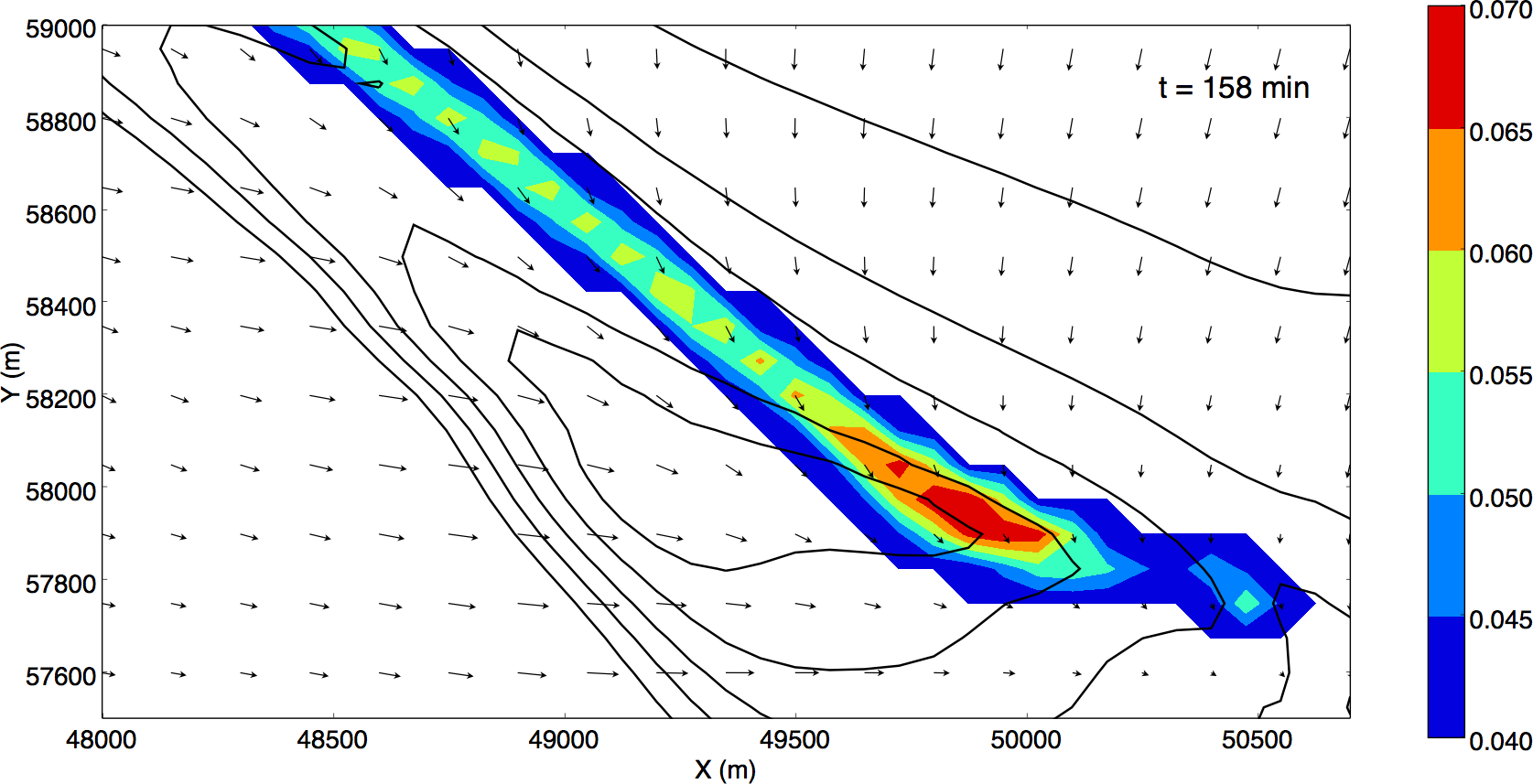}
    \includegraphics[width=0.77\textwidth]{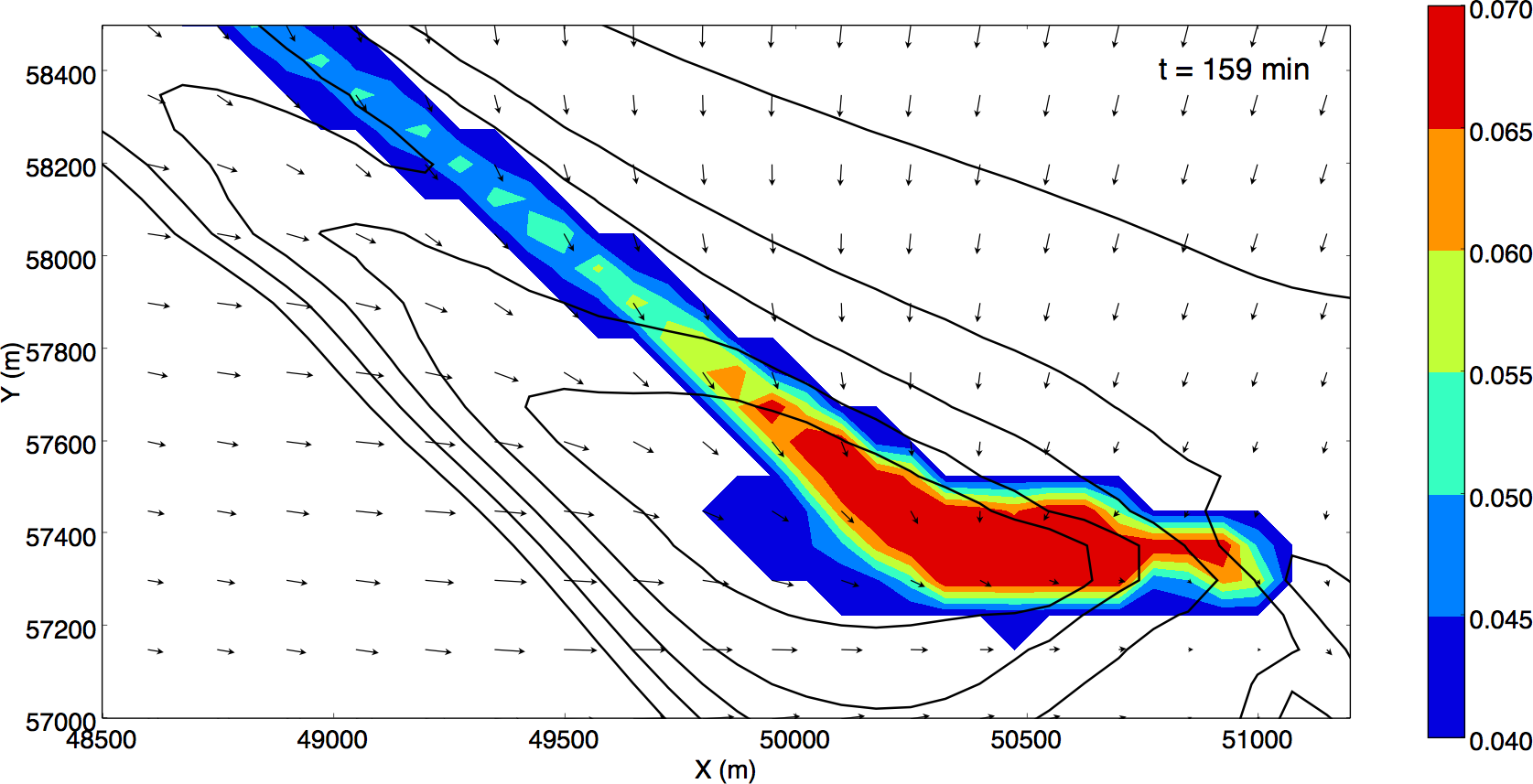}
    \includegraphics[width=0.77\textwidth]{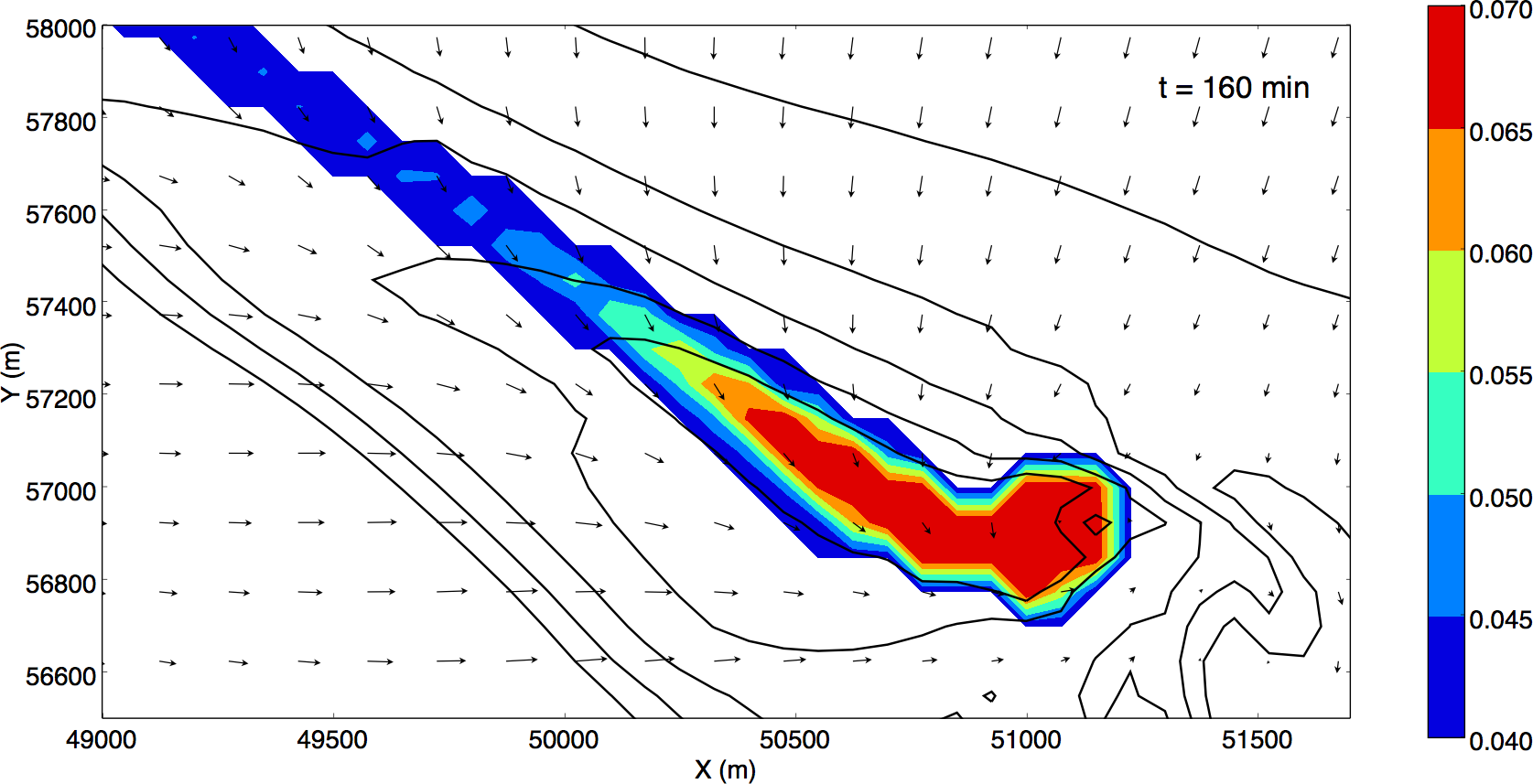}
  \end{center}
  \caption{Vortex sheet roll-up in CM1 simulation at $z=50$ meters and $t=158$, $159$, and $160$ minutes. Vertical vorticity is shaded, vertical velocity is contoured in $1$ m s$^{-1}$ intervals beginning at $1$ m s$^{-1}$, and horizontal wind vectors are plotted as arrows.}
  \label{fig:vortex_image_time_series1.pdf}
\end{figure}

A supercell thunderstorm simulation was investigated to help confirm the conclusions in \cite{cai05} regarding the evolution of mesocyclone vorticity lines prior to and proceeding tornadogenesis. The supercell was simulated using the compressible mode of the nonhydrostatic Bryan Cloud Model 1, CM1 \cite{bryanfritsch02}. The simulation proceeded on a $112.5$ km $\times$ $112.5$ km $\times$ $20.0$ km domain with horizontal grid spacing of $75$ m and vertical grid spacing increasing from $50$ m at the lowest layer to $750$ m at the highest layer. The large and small time steps were $1/4$ s and $1/16$ s, respectively. Typical of idealized storm simulations, a horizontally uniform analytical base state was used (see Figure \ref{skewt-ok-nssl});
terrain, surface fluxes, radiative transfer, and Coriolis acceleration were omitted; and radiative (free slip) lateral (vertical) boundary conditions were imposed. Microphysical processes were parameterized using the double-moment scheme \cite{morrison05}. The subgrid turbulence scheme was similar to that in \cite{deardorff80}. The simulated supercell exhibits features commonly observed in real supercells, including a hook echo reflectivity signature with a cyclonic--anticyclonic vorticity couplet shown in Figure \ref{fig:simulated_features}. 
In Figure~\ref{fig:vortex_image_time_series1.pdf}, a two-minute time interval of the CM1 simulation is shown that illustrates a vortex sheet roll-up prior to tornadogenesis. Notice how the vertical vorticity is concentrating in the area where the eventual tornado is generated; also notice the individual smaller vortices in the rear flank that appear to be ``feeding'' the larger vortex, consistent with the theory proposed in \cite{vortexgas17}.

In order to compute maximum vorticity at different length scales $\varepsilon$, the vorticity field valid on the $75$-m simulation grid was filtered using the Cressman interpolation method \cite{cressman59} with the cutoff radius set to $2\varepsilon$ (consistent with \cite{cai05}). Vorticity lines were then computed near the low-level mesocyclone $\sim500$ m above ground level (AGL) every $5$ minutes once a distinct low-level mesocyclone had formed (as discerned from visual inspection of the $75$-m vorticity field). As in \cite{cai05}, vorticity lines were fit to $300$ m $\leq \varepsilon \leq 9600$ m. Tornadogenesis was considered to occur once the maximum axisymmetric tangential wind velocity, $V_T$, around the intensifying surface vortex associated with the low-level mesocyclone exceeded $20$ m s$^{-1}$. The $V_T$ was retrieved using the vortex detection and characterization technique \cite{potvin13}.

\begin{figure}
  \begin{center}
    \includegraphics[width=0.6\textwidth]{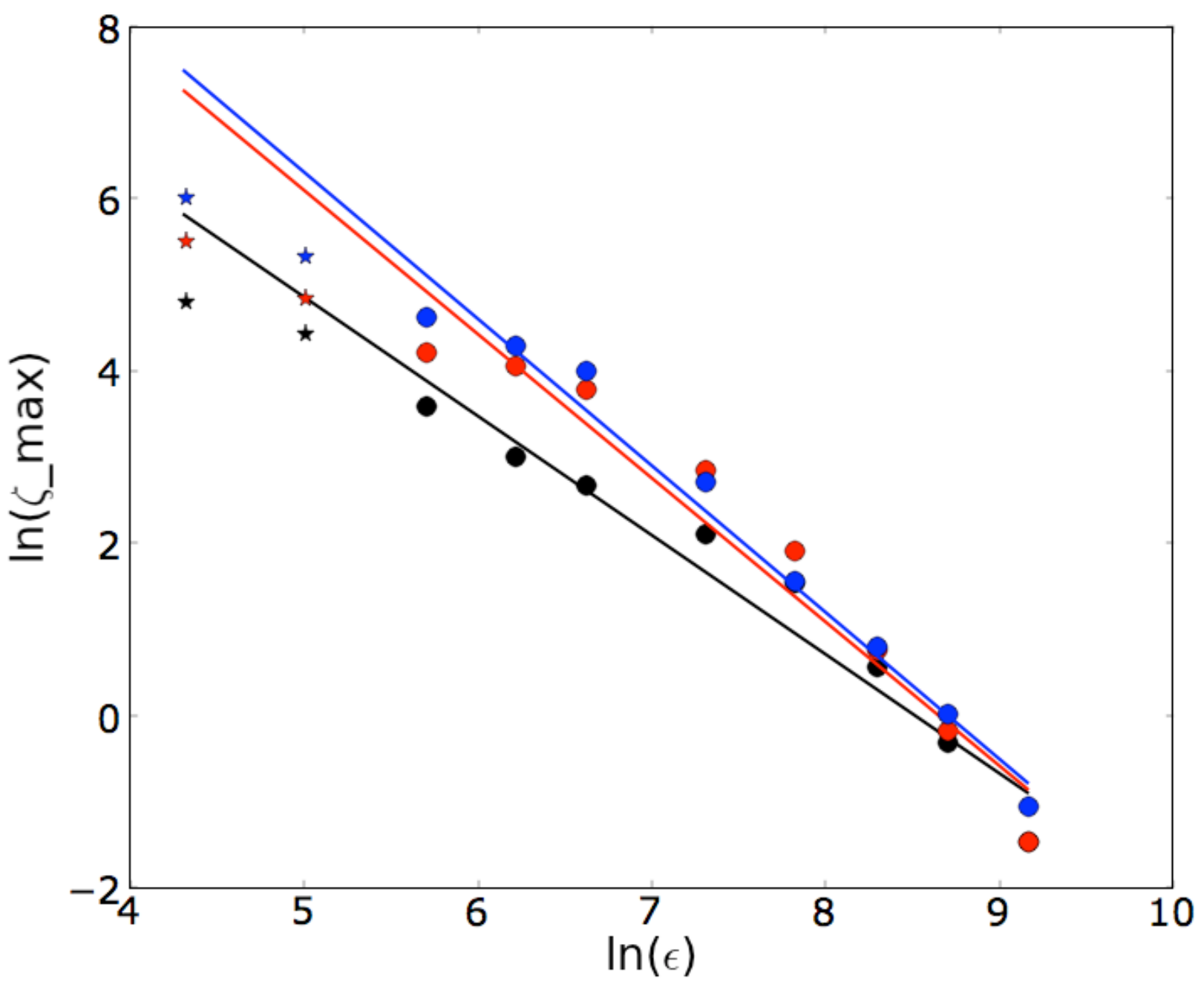}
  \end{center}
  \caption{Vorticity lines computed at $t=134$ min (black, least steep), prior to the development of a discernible surface vortex; at $t=154$ min (red, in-between), by which time a relatively weak surface vortex is present; and at $t=169$ min (blue, steepest), near the time of tornadogenesis. Points used to create the least-squares fit are denoted by dots, while points not used in the vorticity line computation ($\varepsilon<300$ m) are denoted by asterisks.}
  \label{fig:vorticitylines-ok-nssl}
\end{figure}

\begin{figure}
  \begin{center}
    \includegraphics[width=0.6\textwidth]{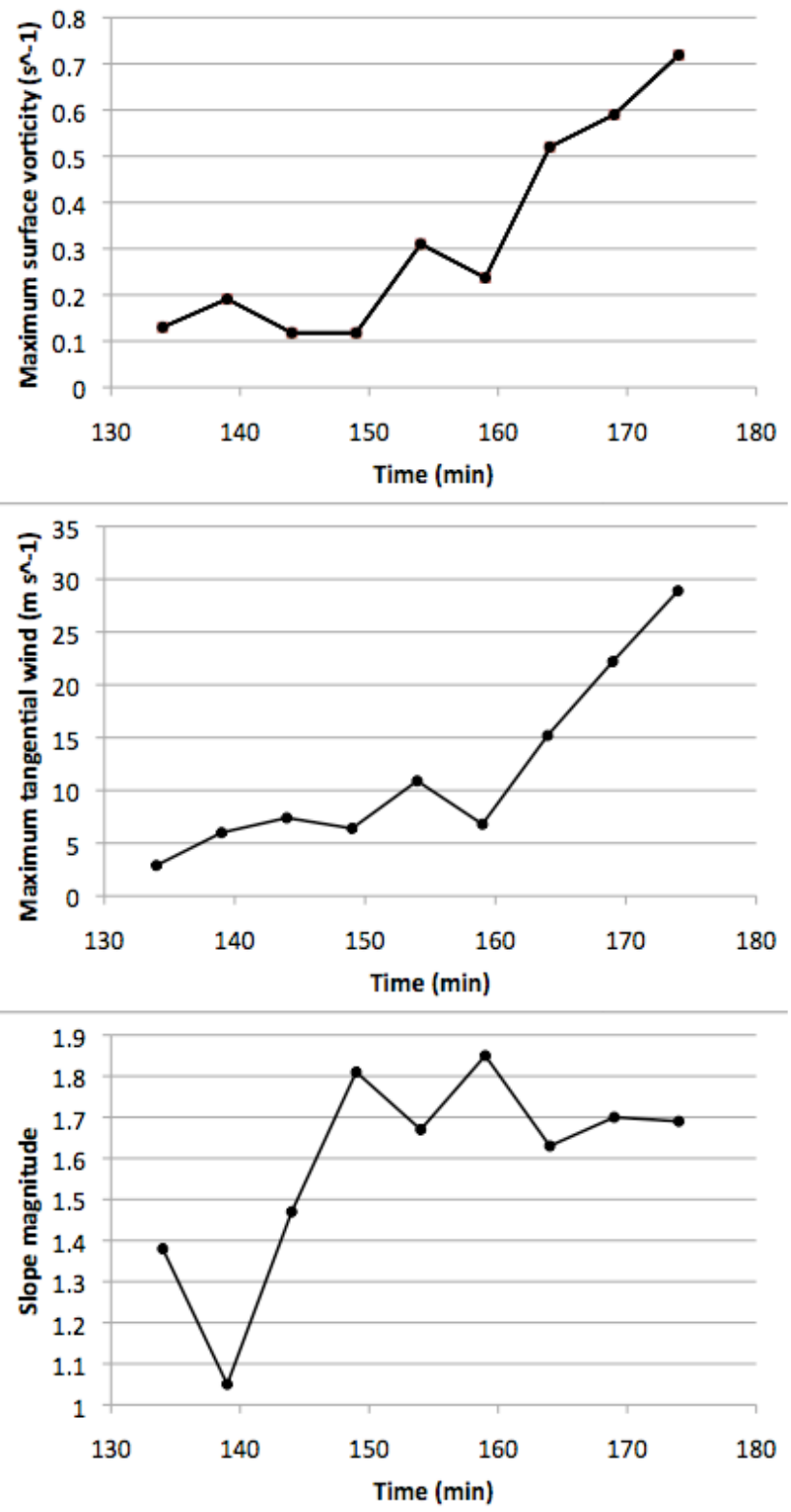}
  \end{center}
  \caption{Time series of (a) maximum surface vorticity (s$^{-1}$), (b) $V_T$ (m s$^{-1}$), and (c) vorticity line slope magnitude.}
  \label{fig:timeseries-ok-nssl}
\end{figure}

As in \cite{cai05}, the vorticity lines steepen prior to tornadogenesis (see Figures \ref{fig:vorticitylines-ok-nssl} and \ref{fig:timeseries-ok-nssl}), consistent with the concentration of vorticity from larger to smaller scales.
In agreement with \cite{cai05}, a power law for vorticity appears to hold for scales exceeding that of the low-level mesocyclone core, but breaks down at smaller scales. In \cite{cai05}, this was attributed to smaller scales being more poorly resolved in the radar dataset. A similar effect occurs in our scenario: the effective model resolution of \cite{frehlich2008} artificially decreases the energy contained at scales approaching the grid spacing. In the absence of positive evidence that the vorticity power law indeed extends to tornadic and smaller scales, we can only offer this as a speculative explanation for the flattening of the vorticity lines at sub-mesocyclone scales.

\section{Power laws in energy spectrum}
\label{sec:energy_spectrum}
We now give an explanation of an increase in vorticity of a developing tornado that is alternative to the power laws of Cai and Wurman. We argue that tornadoes have approximate fractal cross sections and negative temperature (defined as the reciprocal of the rate of change of entropy with respect to energy). We give a power law that relates the increase of the approximate fractal dimension of the cross section of a negative temperature vortex to its energy content. The argument is based on the argument of Chorin.

The well-known power law for dissipation of energy with scale, $\langle E\rangle(k)\sim k^{-5/3}$, where $k$ is the wave number, can be derived by a scaling argument due to Kolmogorov and is also supported by Chorin's filament model using results from a Monte Carlo simulation \cite{chorin-akao91,chorin}.

Chorin's filament model can be applied to analyze a vortex tube in a sparse, homogeneous suspension of tubes. Consider a narrow and straight enough vortex tube $\mathcal{T}$ that has a center line $\mathcal{C}$, parametrized by $s$, and cross sections through $\mathcal{C}(s)$, denoted by $\mathcal{S}(s)$, orthogonal to the center line and such that $\mathcal{S}(s_1)$ and $\mathcal{S}(s_2)$ do not intersect for $s_1\ne s_2$.

Given a point ${\bf x}$ in the vortex tube and $r>0$, we define the ball $B_r({\bf x})=\{{\bf x'}\colon|{\bf x'}-{\bf x}|<r\}$. We take $r$ small enough so that $B_r({\bf x})$ contains no points that belong to other vortex tubes in the suspension. We denote by $\Sigma(s)$ the part of the cross section $\mathcal{S}(s)$ inside $B_r({\bf x})$ and by $\mathcal{C}_r$ the part of the center line of the vortex tube for which $\Sigma(s)$ is non-empty, i.e.,
\begin{equation*}
  \Sigma(s)
  =
  \mathcal{S}(s)\cap B_r({\bf x})
  \quad\text{and}\quad
  \mathcal{C}_r
  =
  \left\{\mathcal{C}(s):\ \Sigma(s)\ne\emptyset\right\}.
\end{equation*}

In order to compute the energy spectrum, Chorin defines, for $r>0$, the vorticity correlation integral
\begin{equation*}
  S_r
  =
  \left\langle
    \int_{\mathcal{T}\cap B_r({\bf x})}\boldsymbol\omega({\bf x})\cdot\boldsymbol\omega({\bf x'})\,d\mathcal{H}_\mathcal{T}
  \right\rangle,
\end{equation*}
where $d\mathcal{H}_\mathcal{T}$ denotes the appropriate Hausdorff measure, related to the set capacity on $\mathcal{T}$, and the average is taken over the ensemble of all possible configurations. Then, using disintegration of measure \cite{schwartz76}, we have
\begin{equation*}
  S_r
  =
  \left\langle
    \boldsymbol\omega({\bf x })\cdot\int_{\mathcal{C}_r}\,ds\int_{\Sigma(s)}\boldsymbol\omega({\bf x'})\,d\mathcal{H}_\Sigma
  \right\rangle.
\end{equation*}
If vorticity is roughly uniform throughout the cross section $\Sigma(s)$ so that $\boldsymbol\omega({\bf x'})\approx\boldsymbol\omega(s)$, and if $|\Sigma|(s)=\mathcal{H}_\Sigma(\Sigma(s))$ denotes the Hausdorff measure of $\Sigma(s)$, then
\begin{equation*}
  S_r
  \approx
  \left\langle
    \boldsymbol\omega({\bf x})\cdot\int_{\mathcal{C}_r}|\Sigma|(s)\boldsymbol\omega(s)\,ds
  \right\rangle.
\end{equation*}
Assuming that the Hausdorff dimension of $\Sigma(s)$ remains constant throughout $\mathcal{T}\cap B_r({\bf x})$ and denoting it by $D_{\Sigma}$, we obtain
\begin{equation*}
  S_r
  =
  \mathcal{O}(r^{D_{\Sigma}+1}).
\end{equation*}
To obtain the vorticity spectrum, $Z(k)$, we integrate the Fourier transform of $S_r$ over a sphere of radius $k=|{\bf k}|$. This gives $Z(k)=\mathcal{O}(k^{-D_{\Sigma}+1})$, and, consequently, the energy spectrum satisfies
\begin{equation}
  \label{eq:e-k-ksq}
  \langle E\rangle(k)
  =
  Z(k)/k^2
  =
  \mathcal{O}(k^{-D_{\Sigma}-1}).
\end{equation}

Let $D_c$ be the dimension of the center line $\mathcal{C}$ of the vortex, $D_\Sigma$ the dimension of the vortex cross section, and $D$ the dimension of the support of the vorticity in the vortex filament $\mathcal{T}$. We assume that the center line of the vortex has Hausdorff dimension one, i.e., $D_c=1$, and therefore \cite{pesin09}
\begin{equation*}
  D
  =
  D_c + D_\Sigma
  =
  1  +  D_\Sigma.
\end{equation*}

\begin{figure}
  \begin{center}
    \includegraphics[width=0.3\textwidth]{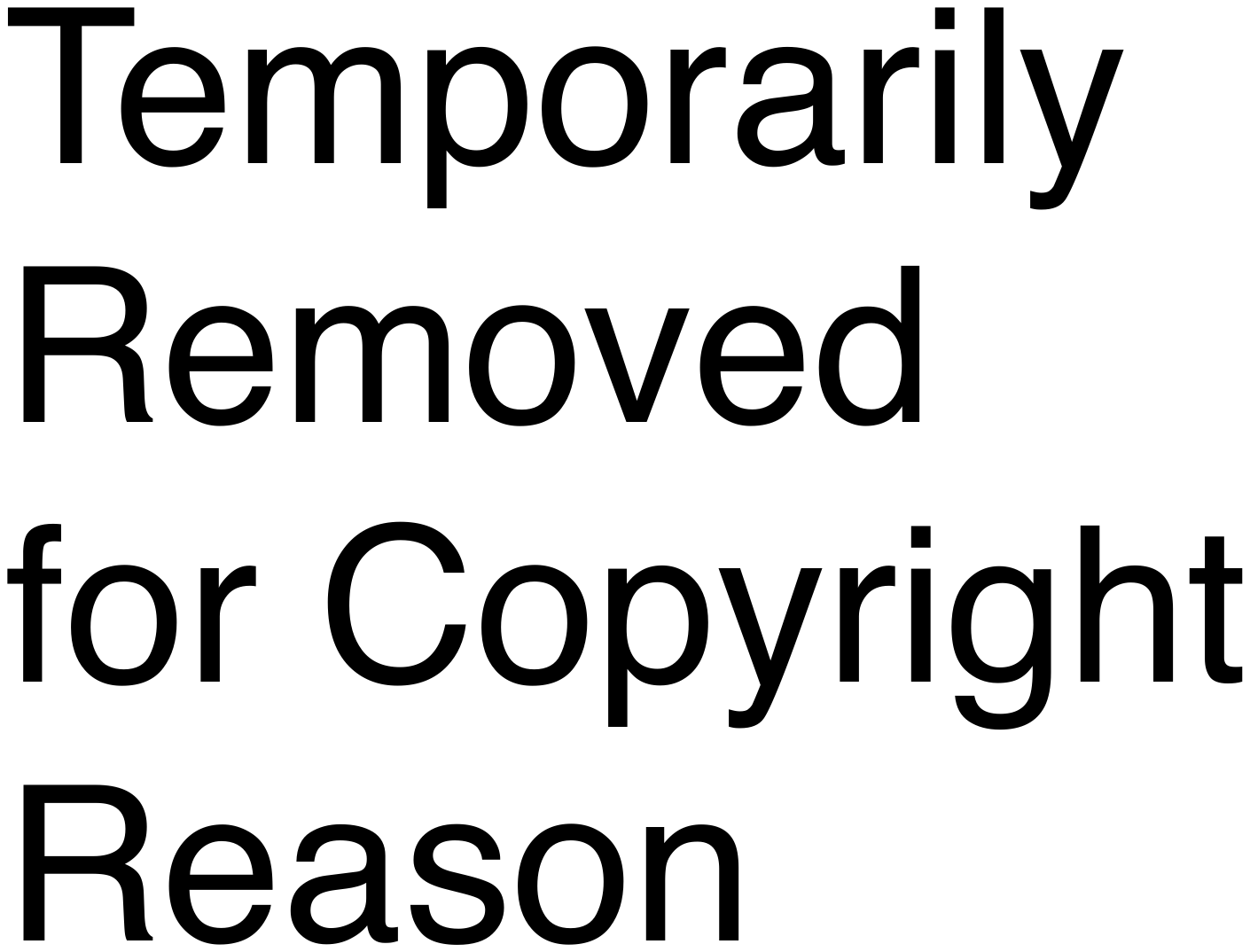}
  \end{center}
  \caption{A tornado and a vortex sheet roll-up over Lake Wilson, Russell County, KS, May 7, 1993; \copyright~Gene Moore. ``Some very subtle features make this one of the most interesting tornado photographs ever taken. As the tornado funnel swept across Lake Wilson, what MAY have been the rear flank downdraft wrapped cyclonically around the south side of the funnel, between the funnel and the camera. Along this shear zone, a train of miniature shear-line vortices developed. Water spray from these 'spin-ups' can be seen about a third of the way across the lake.'' \cite{grazulis97} In light of recent research \cite{orf17}, this feature may also have originated in the forward flank gust front.}
  \label{fig:gene-moore}
\end{figure}

\begin{figure}
  \begin{center}
    \includegraphics[width=\textwidth]{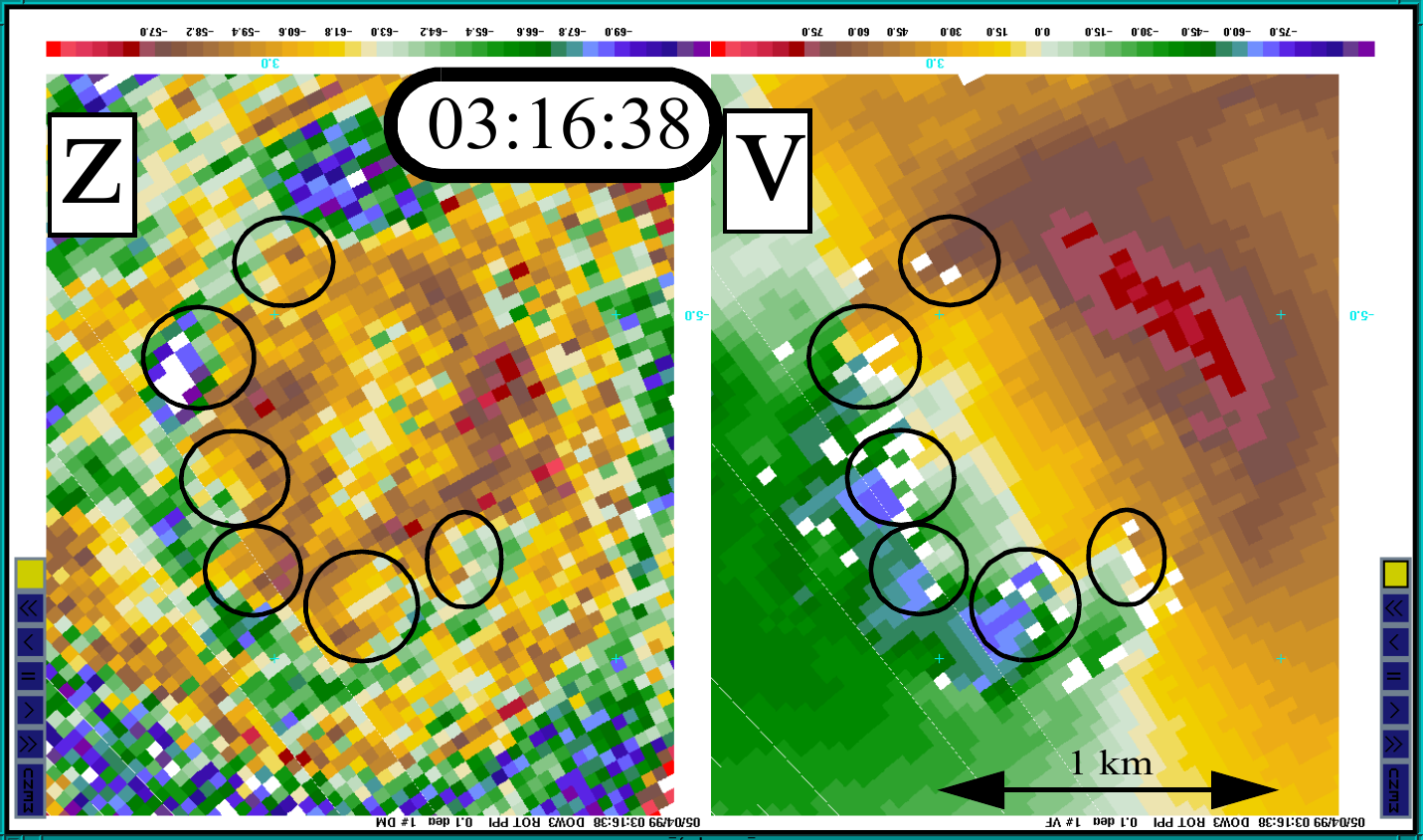}
  \end{center}
  \caption{Reflectivity and velocity radar images from Mulhall, OK, May 3, 1999 tornado \cite{wurman02,kosibawurman08}. A box counting technique can be used to estimate the fractal dimension of the multiple vortex distribution.}
  \label{fig:wurman_radar}
\end{figure}

We suggest that tornadoes have approximately fractal cross sections based on graphical evidence provided in Figures \ref{fig:hier-vort}, \ref{fig:suctionspots}, \ref{fig:tornado_fractal}, and \ref{fig:gene-moore}. This suggestion is further supported by the existence of subvortices within subvortices (see Figure \ref{fig:hier-vort}) as shown in recent videos \cite{wadena10,elrenovideo13}. In particular, a multiple vortex distribution in a larger tornado shown in Figure~\ref{fig:wurman_radar} exhibits an approximate fractal (box counting) dimension in a range from $1.4$ to $1.7$ computed using pairs of grid sizes $\varepsilon$ and $2\varepsilon$ with $\varepsilon$ fine enough to resolve the subvortices in the radar data; the smallest $\varepsilon$ used in the computations was $250$ m. Other radar images in, for example, \cite{wurman02,kosibawurman08} can be analyzed in a similar way.

Therefore, based on the evidence above, we may assume that self-similarity is present throughout a discrete set of scales and thus the cross section of the tornado can be approximated by a fractal. For $ 1<D_\Sigma\le2 $ the energy (see \eqref{eq:e-k-ksq}) satisfies $\langle E\rangle(k)=\mathcal{O}(k^{-\gamma})$ with $2<\gamma\le3$ where $\gamma=D_\Sigma +1$. It follows that for large scales (small $k$), an increase in $\gamma$ in the range from $2$ to $3$ corresponds to an increase of the energy $\langle E\rangle(k)$. This is consistent with the idea that vortices from a vortex sheet feeding a larger tornado cause an increase in the fractal dimension of the tornado's cross-sectional area and an increase in its energy. Thus, an increase in the fractal dimension of the cross-sectional area may be associated with tornadogenesis or strengthening of an existing tornado. Note that this is analogous to Cai's power law, in which a decrease in the (negative) exponent in the power law \eqref{eq:cai_power_law} leads to tornadogenesis or a stronger tornado.

\begin{figure}
  \begin{center}
    \includegraphics[width=0.85\textwidth]{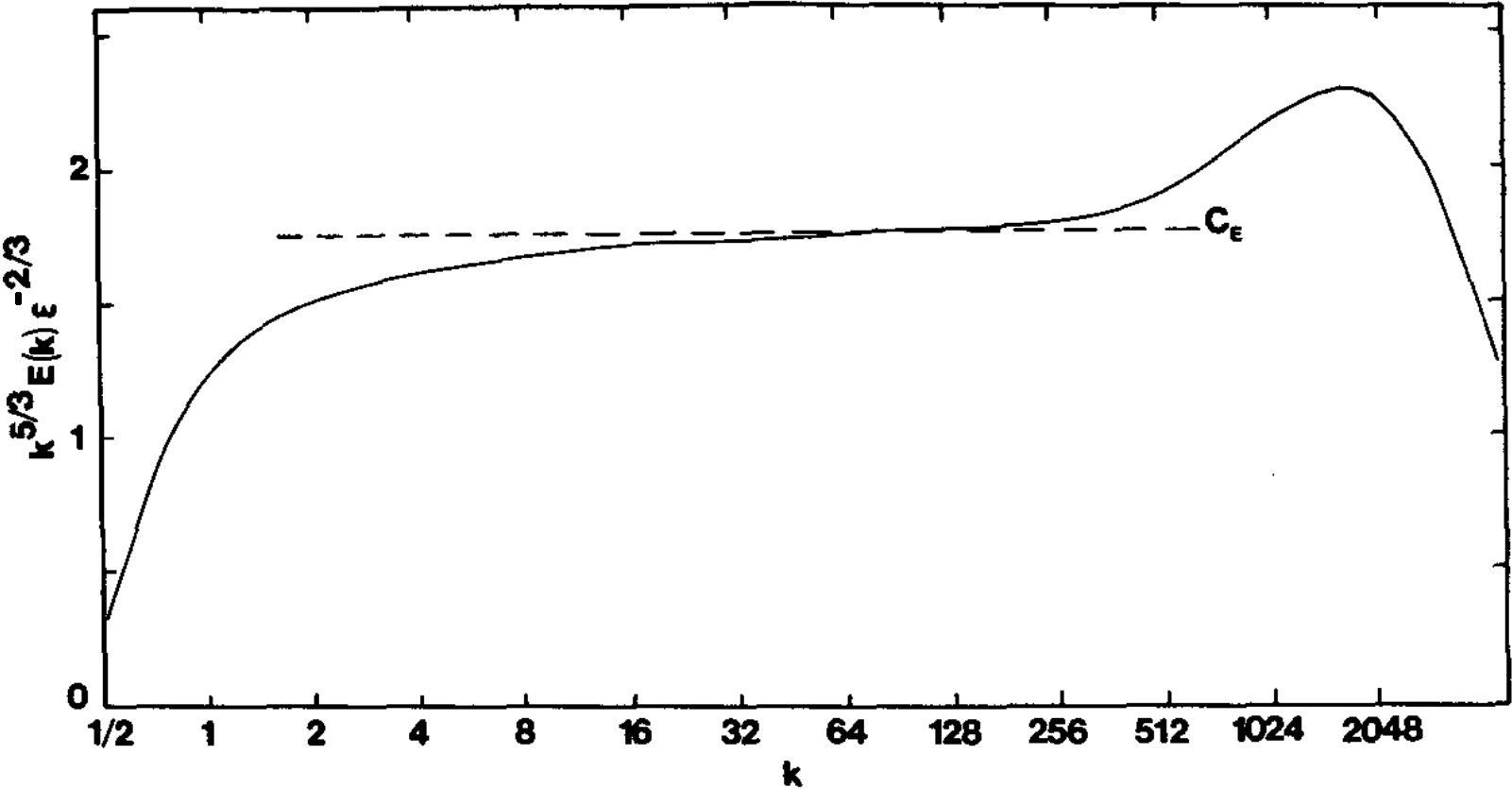}
  \end{center}
  \caption{Inertially weighted energy spectrum corresponding to time $t=8$ for a turbulent flow with low helicity showing high dissipation at large scales (small $k$); \copyright~AMS, \cite{andre77,lilly86b}.}
  \label{fig:lilly-13}
\end{figure}

\begin{figure}
  \begin{center}
    \includegraphics[width=0.85\textwidth]{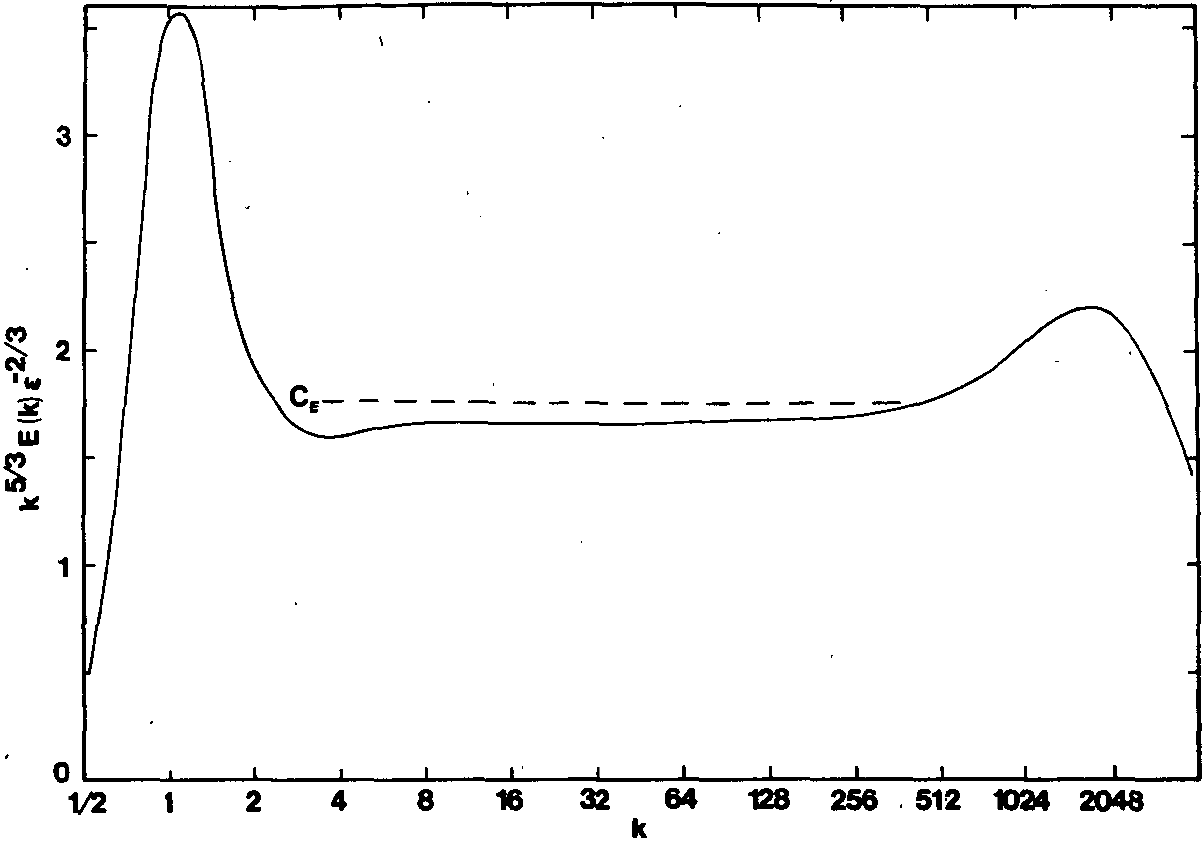}
  \end{center}
  \caption{Inertially weighted energy spectrum corresponding to time $t=12$ for a turbulent flow with high helicity showing low dissipation at large scales (small $k$); \copyright~AMS, \cite{andre77,lilly86b}.}
  \label{fig:lilly-15}
\end{figure}

To gain further insight into the processes that might contribute to tornadogenesis, we consider the effects of helicity in the simplified case of homogeneous isotropic turbulence \cite{lilly86a,lilly86b,andre77}. Writing the Navier--Stokes equation in energy form and taking the Fourier transform, one obtains
\begin{equation*}
  \partial_t\langle E\rangle(k)
  +
  2k^2R^{-1}\langle E\rangle(k)
  =
  Q(k),
\end{equation*}
where $Q(k)$ comes from the nonlinear term in the Navier--Stokes equation \cite{chorin}. This term represents the transfer of energy between wave numbers and has been studied for the case of homogeneous turbulence \cite{waleffe92}. Certain terms have been singled out and studied in relation to inverse energy cascades. These interactions involve three wave numbers. It was found that the net effect of the so-called nonlocal interactions is to transfer energy from intermediate scales to larger scales. These interactions occur between modes with helicity of the same sign. Flows without helicity and dissipation of energy as shown in Figure~\ref{fig:lilly-13}, and flows with helicity and low dissipation of energy as shown in Figure \ref{fig:lilly-15} are discussed in \cite{andre77} and \cite{lilly86b} (the ``epsilon'' on the vertical axes in both Figure \ref{fig:lilly-13} and Figure \ref{fig:lilly-15} stands for the energy dissipation rate per unit mass, not the horizontal grid spacing $\varepsilon$ used in Sections \ref{sec:pseudovorticity}, \ref{sec:cm1}, and \ref{sec:energy_spectrum}). The images represent the results of two numerical experiments and show that isotropic turbulence with helicity inhibits the dissipation of energy at large scales, which is consistent with Figure~\ref{fig:158_zoomed-out}, Figure~\ref{fig:vortex_image_time_series1.pdf}, and Figure~\ref{fig:lilly-15}.

Under the effects of strong rotation, the flow has the tendency to become anisotropic \cite{pouquet10}. Studies have shown that the presence of helicity and low-energy dissipation are unlinked unless the helicity is continuously supplied and/or generated at the energy-containing scales; this is associated with inhomogeneity in the mean field \cite{lilly86a,lilly86b,yokoi93}. Such an inhomogeneity would be supplied by surface friction and the rear flank and forward flank downdrafts and/or their gust fronts (see  Figure~\ref{fig:158_zoomed-out}). The increase in the exponents for the power laws for the vorticity as tornadogenesis approaches (see Figure \ref{fig:timeseries-ok-nssl}) is consistent with the helicity production of the flow at the energy-containing scales (see Figure \ref{fig:lilly-15}). Idealized cross sections of vortices with high helicity exhibit self-similarity (compare Figure \ref{fig:hier-vort} and the image in \cite[p.~168]{arnold99}). Thus the tornado develops at a ``focus'' scale at which much energy and helicity is transferred among other scales. This is consistent with contribution of helicity to the flow \cite{pouquet10}.

\section{Conclusions and Future Work}
\label{sec:conclusions}
In this paper we have reviewed power laws relating various quantities of interest in tornadic fluid flows and have drawn connections between magnitudes of the exponents in the power laws and the intensity of the corresponding tornadic flow. In particular, we have noted the power laws in the tangential component of the velocity as a function of the distance from the axis a tornado, in the vorticity and pseudovorticity as a function of numerical grid spacing in various tornadic and non-tornadic supercells, in the numerically simulated vorticity as a function of grid spacing in a simulation of a tornadogenesis, and in the energy spectrum as a function of the wave number. In all of these cases we have connected the increase in the magnitude of the exponents to the increase in the intensity of the flows and thus associated them with tornadogenesis and maintenance. We have also highlighted the connections between these power laws and the potential fractal-like features of the flows, such as vortices within vortices and fractal cross sections of tornadoes.

Mobile Doppler radar observations of vortices apparently originating in the vortex core suggest that they make a partial revolution about the ambient tornado vortex and then dissipate. In Section \ref{sec:wurman} we discussed the creation of vortices inside and outside the tornado potentially leading to enhancing the tornado's strength. We discussed a particular case in which a power law of the form $v\propto r^{-1}$ was found on one side of an intense tornado and a power law of the form $v\propto r^{-\alpha}$ with $0.5\le\alpha\leq0.6$ on the other \cite{wurman02}. One obvious explanation of such a phenomenon is as follows. Consider a tornado with several vortex sheets spiraling into it from one side of the tornado with zero or minimal vertical vorticity elsewhere. Mathematically, this would correspond to the flow described in \cite{wurman02} with $v$ viewed in a spatially averaged sense. Due to the Kelvin--Helmholtz instability, the vortex sheets would roll up into larger discrete vortices, which, due to their high intensity, would be transient and quickly dissipate, transferring their energy to the ambient flow \cite{vortexgas17}.

While a nonturbulent flow is governed by the Navier--Stokes equations, a turbulent flow corresponds to a large Reynolds number and is then approximately governed by the Euler equations. In the inertial range, a Kolmogorov type law \cite{kolmogorov41} results in an energy cascade from large to small scales, while in the energy-containing range vortices with negative temperatures dissipate and transfer energy to larger scales, generating an inverse energy cascade between scales \cite{vortexgas17}; this process may subsequently lead to an intensification or a genesis of the tornado vortex at the surface. The frequency with which such vortices are produced, their strength, and the stretching of the vortices determine the eventual strength of the tornado. The results of Sections \ref{sec:wurman}, \ref{sec:cm1}, and \ref{sec:energy_spectrum} support this idea. The inverse energy cascade process is related to the concept of negative viscosity (see, e.g., \cite{lilly69,lilly76,bluestein13}). The idea of an inverse energy cascade is supported both by photographic evidence, such as the one in the photo by Gene Moore in Figure~\ref{fig:gene-moore}, as well as by numerical evidence, such as that discussed in Section~\ref{sec:cm1}.

\begin{figure}
  \begin{center}
    \includegraphics[width=0.49\textwidth]{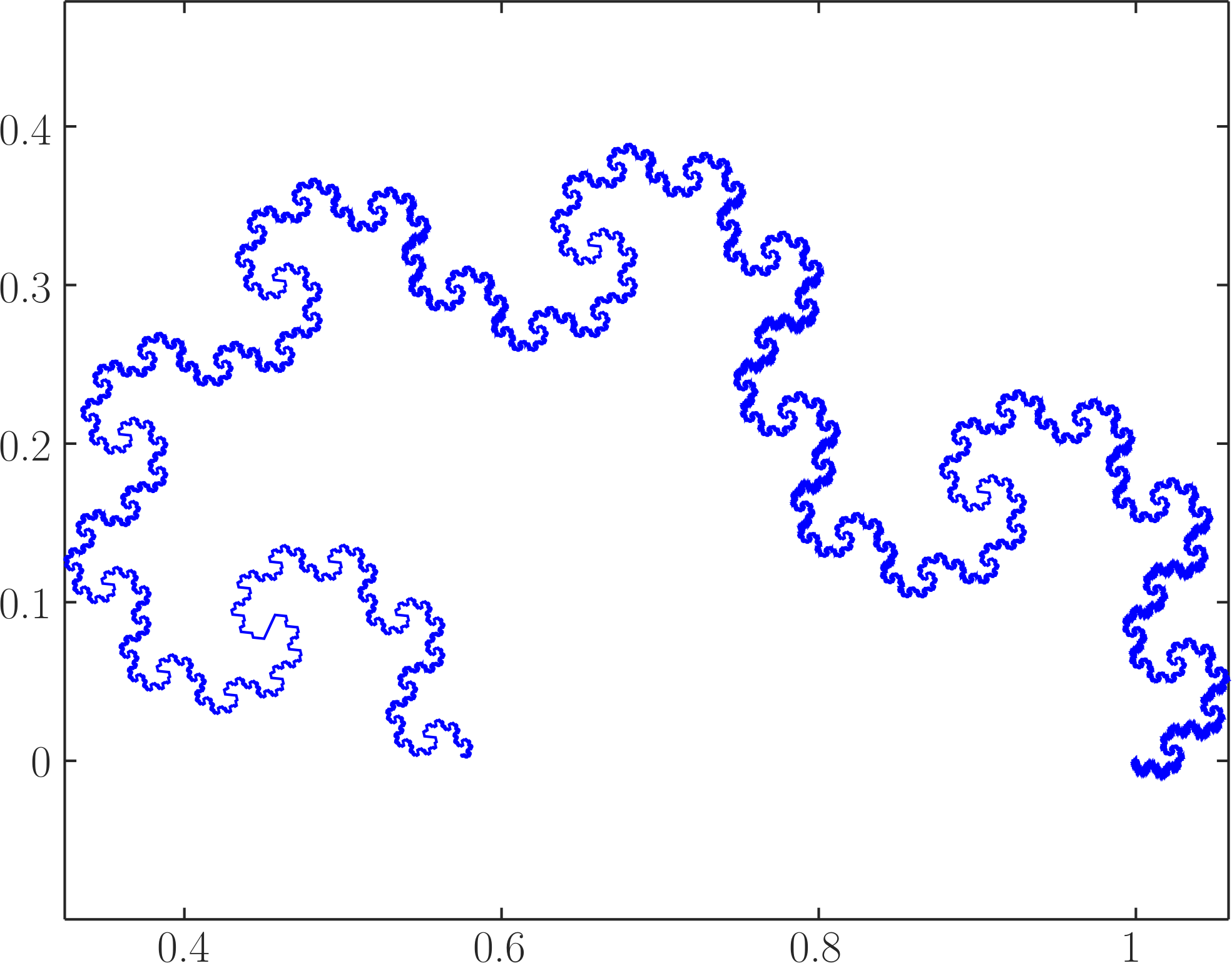}
    \hfill
    \includegraphics[width=0.49\textwidth]{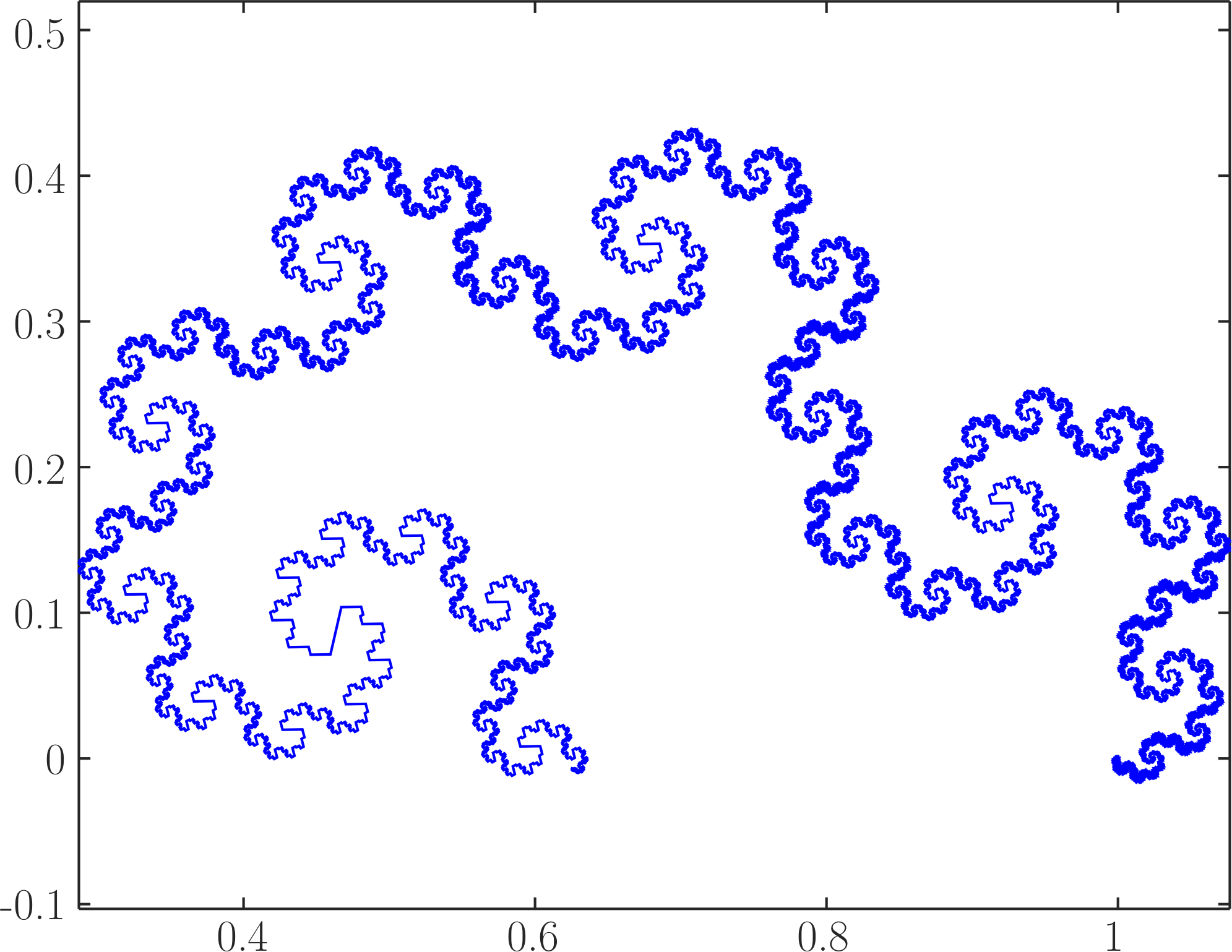}\\[\baselineskip]
    \includegraphics[width=0.49\textwidth]{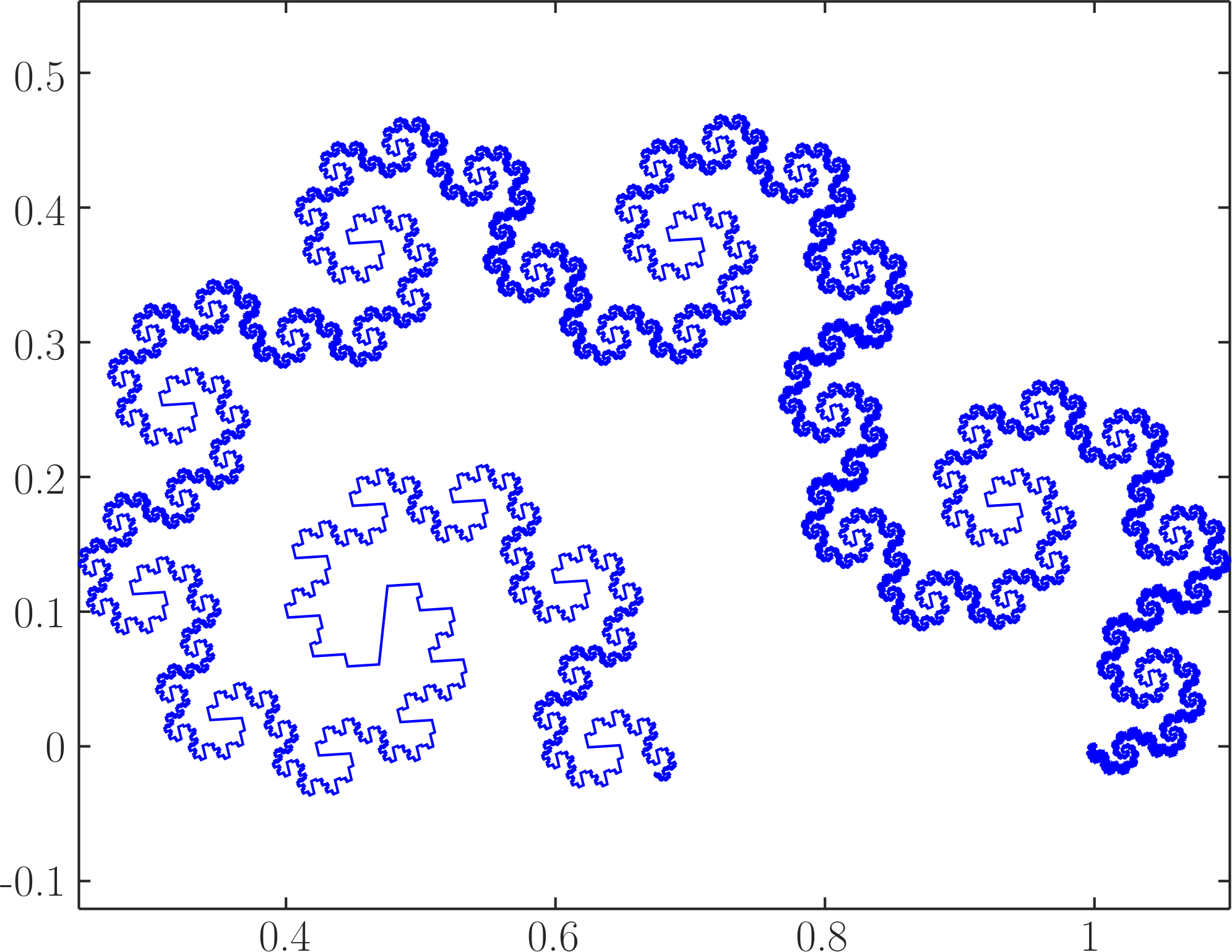}
    \hfill
    \includegraphics[width=0.49\textwidth]{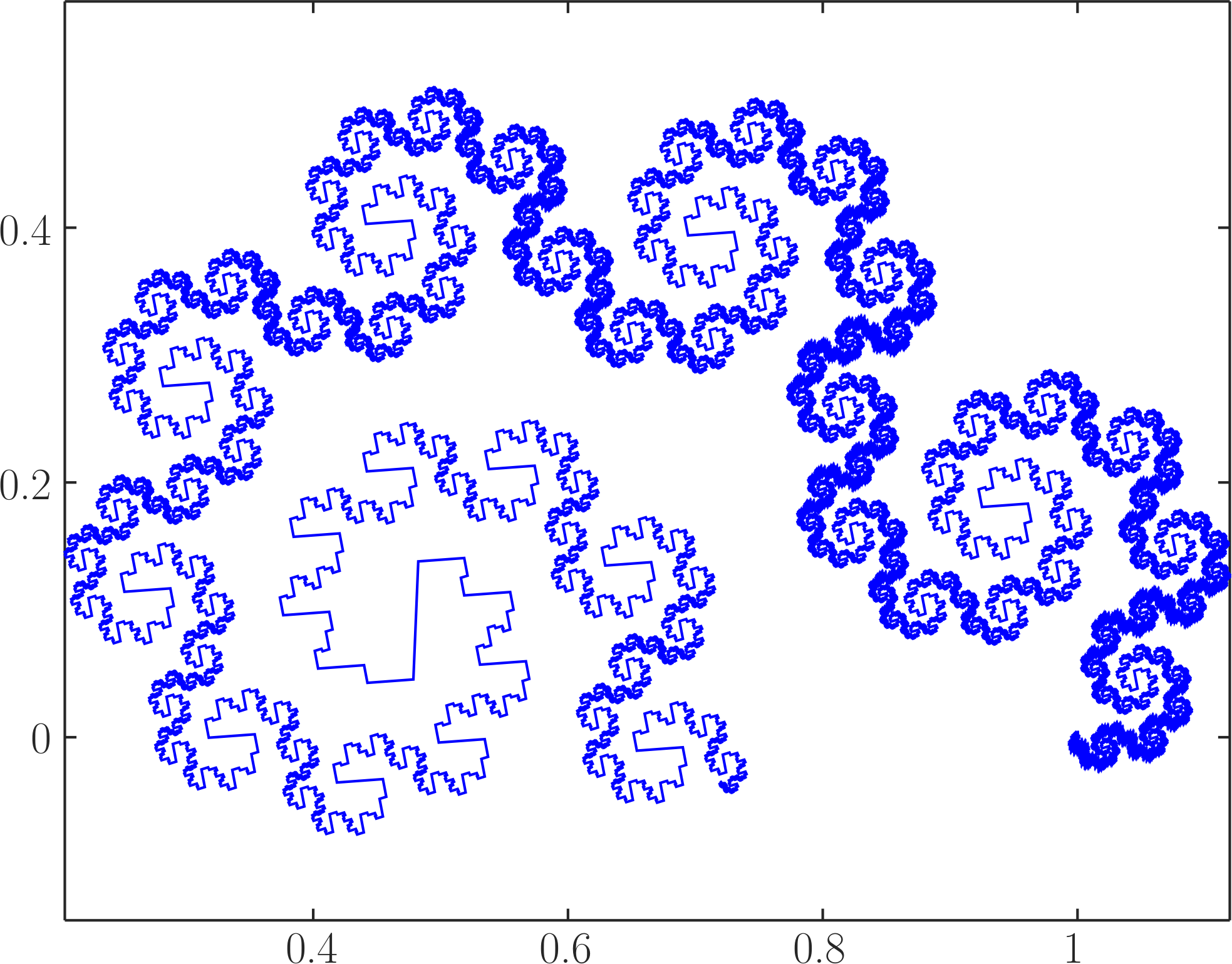}
  \end{center}
  \caption{Examples for ``dragon'' fractals whose fractal dimensions are (left to right, top to bottom) $1.4$, $1.5$, $1.6$, and $1.7$.}
  \label{fig:fractal-curves}
\end{figure}

We have discussed various connections between tornadoes and fractals. As shown in Figure~\ref{fig:hier-vort}, a tornado can be viewed as an intermediate-scale structure in a self-similar cascade of vortex structures exhibiting fractal-like features. The images in Figures~\ref{fig:tornado_fractal}, \ref{fig:gene-moore}, and \ref{fig:wurman_radar}, capturing physical features of particular tornadoes, show vortices within vortices and exhibit features similar to those observed in various known fractals such as the ``dragon'' curves shown in Figure~\ref{fig:fractal-curves}. In addition, since at least two physical scales of self-similarity are present, an approximate fractal dimension can be computed using a box-counting technique. Rough calculations show these fractal dimensions in the range of $1.4$--$1.7$, remarkably consistent with the power law exponents discussed earlier, and also visually consistent with the various dragon fractals with fractal dimensions in similar ranges (see Figure~\ref{fig:fractal-curves}). In Section~\ref{sec:energy_spectrum}, we discussed the possibility of a tornado cross section being a fractal. This would again be consistent with the radar images in Figures~\ref{fig:tornado_fractal} and \ref{fig:wurman_radar}. A possible explanation for these various fractal-like phenomena is readily offered by the process of a roll-up of a vertical vortex sheet. During such roll-ups one would expect a creation of vortices along the vortex sheet \cite{baker90} and further roll-ups of such vortex structures into larger and larger vortices, eventually creating a fractal-like structure. Further thinning of the vortex sheet and intensification of vertical vorticity due to stretching may lead to further creation of smaller, rolled up vortices (in either a direct or inverse cascade), creating additional scales of self-similarity.

Supercell thunderstorms have a quasi-periodic nature, cycling between destructive and rebuilding phases. They typically have a longer life span than generic storms. Curiously, most classic supercells have common features which make them distinctive from other storms of the same scale. These features include, for example, a wall cloud, a tail cloud, and a flanking line. This commonality of features suggests that atmospheric flows that demonstrate themselves as supercells fluctuate near ``attractor'' flows that share some common structure. In the radar reflectivity image of the hook echo region of a supercell thunderstorm shown in Figure \ref{fig:tornado_fractal}, the hooks on the boundary of the region represent successive vortices in a vortex sheet. Such vortices provide periodic pulses of energy to the tornado. An analogy can thus be drawn with the work of Kuznetsov \cite{kuznetsov11}, in which periodic pulses introduced into a dynamical system lead to a Smale--Williams attractor. Sasaki describes the ``wrap-around mechanism,'' a nonlinear process related to his proposed entropic balance theory, used to explain tornadogenesis and features of both a tornado and its parent tornadic supercell~\cite{sasaki14}. He relates the wrap-around mechanism to a nonlinear attractor. We would like to pursue this idea for further study. 

Based on the discussion in Section~\ref{sec:energy_spectrum}, we plan further exploration of the relationships among helicity, temperature (in the vortex gas sense), self-similarity, and the power laws discussed in this paper. The resulting benefits to our understanding of helical atmospheric vortices could improve operational tornadic prediction. To the degree that the vorticity power law extends from observable to tornadic scales, it may also be possible to improve tornado detection and perhaps even estimate maximum tangential winds in tornadoes, as discussed in \cite{cai05}. These hypotheses should be tested using real radar observations of tornadic and non-tornadic supercells.


\bibliography{tornado}
\bibliographystyle{abbrv}

\end{document}